\documentstyle[12pt]{article}
\def\CH{{\cal H}}
\def\CBPi{\bar{\cal P}{}^\prime}
\def\CBPii{\bar{\cal P}{}^{\prime\prime}}
\def\CBP{\bar{\cal P}}
\def\P{{\cal P}}

\begin{document}

\thispagestyle{empty}
\begin{flushright}
                                                        FIAN/TD/98-12
\end{flushright}

\vspace{1.5cm}
\begin{center}
{\Large EXISTENCE THEOREM}
\end{center}

\begin{center}
{\Large FOR SPLIT INVOLUTION CONSTRAINT ALGEBRA}
\end{center}

\vspace{0.5cm}

\begin{center}
{\large I.A. Batalin$^{*}$ }  
\end{center}

\begin{center}
{\it I.E. Tamm Theory Division, \\
P.N. Lebedev Physics Institute, Russian Academy of Sciences, \\
53 Leninsky Prospect, 117924  Moscow, Russia }
\end{center}

\vspace{0.5cm}

\begin{center}
{\large S.L. Lyakhovich$^{**}$ }        
\end{center}

\begin{center}{\it Theoretical Physics Department, \\
Tomsk State University, 634050  Tomsk, Russia}
\end{center}

\vspace{0.5cm}

\begin{center}
{\large I.V. Tyutin$^{***}$ }   
\end{center}

\begin{center}
{\it I.E. Tamm Theory Division, \\
P.N. Lebedev Physics Institute, Russian Academy of Sciences, \\
53 Leninsky Prospect, 117924  Moscow, Russia }
\end{center}

\vspace{0.5cm}
\begin{center} ABSTRACT \end{center}
\vspace{0.5cm}

{\rm Existence theorem is proven for the generating equations of the split
involution constraint algebra. The structure of the general solution is established,
and the characteristic arbitrariness in generating functions is described. }

\vfill
\noindent
$^*$ E-mail address: batalin@lpi.ac.ru

\noindent
$^{**}$ E-mail address: sll@phys.tsu.ru

\noindent
$^{***}$ E-mail address: tyutin@lpi.ac.ru

\newpage

\section {Introduction}

In previous papers [1,2] of the present authors a special class of
dynamical systems with first- and second-class constraints has been
considered whose characteristic feature was an interesting algebraic
structure generated by the constraints, which was called "split
involution". In fact, the split involution means that the second--class
constraints are Sp(2)--polarized which
is formally similar to the global Sp(2)--invariant structure of the
ghost--antighost symmetric quantization of general gauge theories [3,4].
It appeared to be possible in this case to quantize the
dynamical systems by applying a generalized version of the BFV
formalism, without explicit use of the Dirac brackets despite the
presence of second--class constraints. In papers [1,2] the generating
equations were formulated for the nilpotent generating operators
determining physical sector, and for the unitarizing Hamiltonian as well,
and the natural automorphisms of a solution to these equations were also
found which the physical sector does not depend on. In the present
paper we prove in details the existence of a solution to the
generating equations, and then describe the characteristic
arbitrariness of this solution. The proof is based on complete
abelianization of the split involution algebra with the help of the
admissible transformations (i.e. the ones which can be extended to
become a natural automorphism of the generating equations).

{\bf Notations and conventions.}
In what follows $A^{[\mu\nu\lambda\ldots]}$ means a matrix antisymmetric
in the indices inside the square brackets.
$C_{\mu|(\nu)_k(\lambda)_l|(\sigma)_m}$ and
$s_{\mu|(\nu)_k(\lambda)_l}$  are abbreviated notations for the tensors
$C_{\mu|\nu_1\ldots\nu_k\lambda_{k+1}\ldots\lambda_{k+l}|
\sigma_1\ldots\sigma_m}$ and
$s_{\mu|\nu_1\ldots\nu_k\lambda_{k+1}\ldots\lambda_{k+l}}$ symmetric in
the indices not divided by vertical line. By $s(x,y)$ we denote a function
of the type
$$
s(x,y)=\sum_{m,n}s_{(\mu)_m(\nu)_n}(x_\mu)^m(y_\nu)^n,
$$
$$
(x_\mu)^m\equiv x_{\mu_1}\ldots x_{\mu_m}.
$$
All the formulae mentioned in the text are enumerated independently within
each separate Section. When reading through a given Section, the reference (n)
means the n-th  mentioned formula of this Section. When mentioning the m-th
formula of another, say k-th Section, we use the reference (k.m).

\section {Generating equations}

Let
$$
\xi=(q^i, p_i), i=1,\ldots, n=n_+ + n_ - \, ,\quad
(q^i)^\dagger=q^i,\quad(p_i)\dagger=p_i(-1)^{\varepsilon_i},
$$
$$
\varepsilon(q^i)=\varepsilon(p_i)\equiv\varepsilon_i,\quad\hbox{gh}^\prime(q^i)=
\hbox{gh}^{\prime\prime}(q^i)=-\hbox{gh}^\prime(p_i)=
-\hbox{gh}^{\prime\prime}(p_i)=0,
$$
be a set of the original phase variable operators whose equal time nonzero
supercommutation relations are
$$
(\imath\hbar)^{-1}[q^i,p_j]=\delta^i_j.
$$
Further, let us introduce the Hamiltonian,
$$
H=H(p,q), \quad\varepsilon(H)=0,
$$
and the constraint operators,
$$
T_\alpha=T_\alpha(p,q),\quad\varepsilon(T_\alpha)\equiv
\varepsilon_\alpha, \quad
\alpha=1,\ldots,m^\prime=m^\prime_+ + m^\prime_-,
$$

$$
T^a_\mu=T^a_\mu(p,q),\quad\varepsilon(T^a_\mu)\equiv\varepsilon_\mu,\quad
a=1,2,
$$
$$
\mu=1,\ldots,m^{\prime\prime}=m^{\prime\prime}_++m^{\prime\prime}_-,\quad
m^{\prime\prime}_-=2k,\quad
m_\pm\equiv m^\prime_\pm+m^{\prime\prime}_\pm<n_\pm.
$$

As a next step, let us introduce the ghost phase variable operators. We assign
a ghost canonical pair to each first class constraint operator:
$$
T_\alpha\quad\rightarrow\quad(C^{\prime\alpha},\CBPi_\alpha),\quad
\alpha=1,\ldots,m^\prime \, ,
$$
$$
\varepsilon(C^{\prime\alpha})=\varepsilon(\CBPi_\alpha)=
\tilde{\varepsilon}_\alpha+1 ,\quad
\hbox{gh}^\prime(C^{\prime\alpha})=-\hbox{gh}^\prime(\CBPi_\alpha)=1,\quad
\hbox{gh}^{\prime\prime}(C^{\prime\alpha})=\hbox{gh}^{\prime\prime}
(\CBPi_\alpha)=0,
$$
$$
(C^{\prime\alpha})^\dagger=C^{\prime\alpha},\quad(\CBPi_\alpha)^\dagger=
-\CBPi_\alpha(-1)^{\tilde{\varepsilon}_\alpha}.
$$
In the same way we assign a ghost canonical pair to each (a = 1,2) pair of the
second class constraint operators:
$$
T^a_\mu\quad\rightarrow\quad(C^{\prime\prime\mu},\CBPii_\mu),\quad
\mu=1,\ldots,m^{\prime\prime},
$$
$$
\varepsilon(C^{\prime\prime\mu})=\varepsilon(\CBPii_\mu)=\varepsilon_\mu+1,
\quad
\hbox{gh}^\prime(C^{\prime\prime\mu})=\hbox{gh}^\prime(\CBPii_\mu)=0,\quad
\hbox{gh}^{\prime\prime}(C^{\prime\prime\mu})=
-\hbox{gh}^{\prime\prime}(\CBPii_\mu)=1,
$$
$$
(C^{\prime\prime\mu})^\dagger=C^{\prime\prime\mu},\quad
(\CBPii_\mu)^\dagger=-\CBPii_\mu(-1)^{\varepsilon_\mu}.
$$
The equal time nonzero supercommutators of the ghost operators introduced are
$$
(\imath\hbar)^{-1}[C^{\prime\alpha},\CBPi_\beta]=\delta^\alpha_\beta,\quad
(\imath\hbar)^{-1}[C^{\prime\prime\mu},\CBPii_\nu]=\delta^\mu_\nu.
$$
Now, we introduce the generating operators

$$
\Omega^a(q,p,C^\prime,\CBPi,C^{\prime\prime},\CBPii),\quad
\varepsilon(\Omega^a)=1,\quad
\hbox{gh}^\prime(\Omega^a)=0,\quad\hbox{gh}^{\prime\prime}(\Omega^a)=1,
$$

$$
\Omega(q,p,C^\prime,\CBPi,C^{\prime\prime},\CBPii),\quad\varepsilon(\Omega)=1,
\quad\hbox{gh}^{\prime}(\Omega) =1,\quad\hbox{gh}^{\prime\prime}(\Omega)=0,
$$

$$
K(q,p,C^\prime,\CBPi,C^{\prime\prime},\CBPii),\quad\varepsilon(K)=0,\quad
\hbox{gh}^\prime(K)=2,\quad\hbox{gh}^{\prime\prime}(K)=-2,
$$

$$
\CH(q,p,C^\prime,\CBPi,C^{\prime\prime},\CBPii),\quad\varepsilon(\CH)=0,\quad
\hbox{gh}^\prime(\CH)=0,\quad\hbox{gh}^{\prime\prime}(\CH)=0,
$$

$$
\Lambda(q,p,C^\prime,\CBPi,C^{\prime\prime},\CBPii),\quad
\varepsilon(\Lambda)=1,\quad
\quad\hbox{gh}^\prime(\Lambda)=1,\quad
\hbox{gh}^{\prime\prime}(\Lambda)=-2,
$$
to satisfy the set of generating equations
$$
\begin{array}{c}
[\Omega^a,\Omega^b]=0  , \quad (\Omega^a)^\dagger = \Omega^a,  \\[8pt]
{[}\Omega^a  , \,\Omega \, ] = 0  ,\quad  (\Omega \, )^\dagger = \Omega \, ,
\\[8pt]
[ \Omega \, , \,\Omega \, ] =
\varepsilon_{ab}(\imath\hbar)^{-1}[\Omega^b,[\Omega^a,K]],
\, \, \,
(K)^\dagger = K, \\[8pt]
[\Omega^a,\CH] = 0 \ , \quad  (\CH)^\dagger = \CH, \\[8pt]
[\Omega,\CH \, ] = \varepsilon_{ab}
(\imath\hbar)^{-1}[\Omega^b,[\Omega^a,\Lambda]]\, , \, \,
 (\Lambda)^\dagger = \Lambda.
\end{array}
\eqno{(1)}
$$
These equations admit the following group of natural automorphisms
$$
A=A_{1} \, {\times} \,  A_{2} \, {\times} \,  A_{3}
\eqno{(2)}
$$
where $A_{1}$ is the standard unitary group
$$
\Omega^a\quad\rightarrow\quad U^{-1}\Omega^aU,
$$
$$
\Omega\quad\rightarrow\quad U^{-1}\Omega U,\quad K\quad\rightarrow\quad
U^{-1}KU,
$$
$$
\CH\quad\rightarrow\quad U^{-1}\CH U,\quad\Lambda\quad\rightarrow\quad
U^{-1}\Lambda U,
$$
$A_{2} = GL(2,R)$ is the group of $c$-numerical nondegenerate transformations
$$
\Omega^a\quad\rightarrow \quad S^a_b \, \Omega^b,\quad
\Omega\quad\rightarrow\quad\Omega,
\quad\CH\quad\rightarrow\quad\CH,
$$
$$
K\quad\rightarrow\quad\lambda^{-1}K,\quad\Lambda\quad\rightarrow\quad
\lambda^{-1} \Lambda ,\quad \lambda\equiv\hbox{det}(S^a_b),
$$
and $A_{3}$ is the group of exact shifts

$$
\Omega^a\quad\rightarrow\quad\Omega^a,
$$

$$
\Omega\quad\rightarrow\quad\Omega+\varepsilon_{ab}(\imath\hbar)^{-2}[\Omega^b,
[\Omega^a,\Xi]],
$$

$$
K\quad\rightarrow\quad K+2(\imath\hbar)^{-1}[\Omega,\Xi]+\varepsilon_{ab}
(\imath\hbar)^{-3}[[\Omega^b,\Xi],[\Omega^a,\Xi]]+(\imath\hbar)^{-1}[\Omega^a,
X_a],
$$

$$
\CH\quad\rightarrow\quad\CH+(\imath\hbar)^{-1}[\Omega,\Psi]+\varepsilon_{ab}
(\imath\hbar)^{-1}[\Omega^b,[\Omega^a,\Phi]],\quad[\Omega^a,\Psi]=0,
$$

$$\begin{array}{c}
\Lambda\quad\rightarrow\quad\Lambda+(\imath\hbar)^{-1}[\Xi,\CH]+
(\imath\hbar)^{-1}[\Omega,\Phi]+{1\over2}(\imath\hbar)^{-1}[K,\Psi]+\\[9pt]
+(\imath\hbar)^{-2}[\Xi,[\Omega,\Psi]]+\varepsilon_{ab}(\imath\hbar)^{-3}
[[\Xi,\Omega^b],[\Omega^a,\Phi]]+(\imath\hbar)^{-1}[\Omega^a,Y_a].
\end{array}
$$

In the present paper we prove the existence of a solution to the classical
counterpart of the generating equations (2.1). In classical approximation,
the symbols of the generating operators (i.e. the generating functions)
should satisfy the classical equations (2.1) (in what follows all the
quantities are classical ones) with the changes
$(\imath\hbar)^{-1}[\, ,\,]$ $\to$ $\{ \, , \, \}$:
$$
\begin{array}{c}
{\{ }\Omega^a,\Omega^b{ \} }=0  , \quad (\Omega^a)^* = \Omega^a,  \\[8pt]
{{\{ }}\Omega^a  , \,\Omega \, { \} } = 0  ,\quad  (\Omega \, )^* =
\Omega \, , \\[8pt]
{\{ } \Omega \, , \,\Omega \, { \} } =
\varepsilon_{ab}{\{ }\Omega^b,{\{ }\Omega^a,K{ \} }{ \} },
\, \, \,
(K)^* = K, \\[8pt]
{\{ }\Omega^a,\CH{ \} } = 0 \ , \quad  (\CH)^* = \CH, \\[8pt]
{\{ }\Omega,\CH \, { \} } = \varepsilon_{ab}
{\{ }\Omega^b,{\{ }\Omega^a,\Lambda{ \} }{ \} }\, , \, \,
 (\Lambda)^* = \Lambda.
\end{array}
\eqno{(3)}
$$
Let us write down the lowest order contribution to the ghost power series
expansion of generating functions
$$
\begin{array}{c}
\Omega^a=C^{\prime\prime\mu}T^a_\mu+{1\over2}C^{\prime\prime\nu}
C^{\prime\prime\mu}U^{a\rho}_{\mu\nu}\CBPii_\rho(-1)^{\varepsilon_\nu+
\varepsilon_\rho}+C^{\prime\alpha}C^{\prime\prime\mu}
\tilde{U}^{a\beta}_{\mu\alpha}\CBPi_\beta(-1)^{\tilde{\varepsilon}_\alpha+
\tilde{\varepsilon}_\beta}+ \\[7pt]
{1\over2}(-1)^{(\varepsilon_\nu+\varepsilon_\mu\tilde{\varepsilon}_\alpha)}
C^{\prime\alpha}C^{\prime\prime\nu}C^{\prime\prime\mu}
\tilde{U}^{a\beta\rho}_{\mu\nu\alpha}\CBPii_\rho\CBPi_\beta
(-1)^{\varepsilon_\rho} + \ldots \, ,
\end{array}
$$
$$
\Omega=C^{\prime\alpha}T_\alpha+{1\over2}C^{\prime\beta}
C^{\prime\alpha}\tilde{U}^\gamma_{\alpha\beta}\CBPi_\gamma
(-1)^{\tilde{\varepsilon}_\beta+\tilde{\varepsilon}_\gamma}+
C^{\prime\alpha}C^{\prime\prime\mu}U^\nu_{\mu\alpha}\CBPii_\nu
(-1)^{\tilde{\varepsilon}_\alpha+\varepsilon_\nu}+\ldots,
$$

$$
K={1\over2}C^{\prime\beta}C^{\prime\alpha}W^{\mu\nu}_{\alpha\beta}
\CBPii_\nu\CBPii_\mu(-1)^{\tilde{\varepsilon}_\beta+\varepsilon_\nu}+\ldots,
$$

$$
\CH=H-C^{\prime\prime\mu}V^\nu_\mu\CBPii_\nu(-1)^{\varepsilon_\nu}-
C^{\prime\alpha}\tilde{V}^\beta_\alpha\CBPi_\beta
(-1)^{\tilde{\varepsilon}_\beta}+\ldots,
$$

$$
\Lambda={1\over2}C^{\prime\alpha}W^{\mu\nu}_\alpha\CBPii_\nu\CBPii_\mu
(-1)^{\varepsilon_\nu}+\ldots,
$$
where the coefficient functions
$$
U^{a\rho}_{\mu\nu},\quad\tilde{U}^{a\beta}_{\mu\alpha},
\quad
U^\nu_{\mu\alpha},\quad \tilde{U}^\gamma_{\alpha\beta},
\quad
\tilde{U}^{a\beta\rho}_{\mu\nu\alpha}
\quad
W^{\mu\nu}_{\alpha\beta},
\quad
W^{\mu\nu}_\alpha \, ,
\quad V^\nu_\mu,\quad \tilde{V}^\beta_\alpha \, ,
\eqno{(4)}
$$
are supposed to carry the natural parity
and to have the following symmetry properties in their indices
$$
U^{a\rho}_{\mu\nu}=-U^{a\rho}_{\nu\mu}(-1)^{\varepsilon_\mu\varepsilon_\nu},
\quad
\tilde{U}^\gamma_{\alpha\beta}=-\tilde{U}^\gamma_{\beta\alpha}
(-1)^{\tilde{\varepsilon}_\alpha\tilde{\varepsilon}_\beta} \, ,
\quad
\tilde{U}^{a\gamma\rho}_{\mu\nu\alpha}=-\tilde{U}^{a\gamma\rho}_{\nu\mu\alpha}
(-1)^{\varepsilon_\mu\varepsilon_\nu+\varepsilon_\nu
\tilde{\varepsilon}_\alpha+\tilde{\varepsilon}_\alpha\varepsilon_\mu},
$$
$$
W^{\mu\nu}_{\alpha\beta}=-W^{\nu\mu}_{\alpha\beta}
(-1)^{\varepsilon_\mu\varepsilon_\nu}=-W^{\mu\nu}_{\beta\alpha}
(-1)^{\tilde{\varepsilon}_\alpha\tilde{\varepsilon}_\beta},
\quad
W^{\mu\nu}_\alpha=-W^{\nu\mu}_\alpha(-1)^{\varepsilon_\mu\varepsilon_\nu}.
$$
It follows from the generating equations (2.1), (2.3) that the constraints and
structure functions should satisfy the algebra
$$
\begin{array}{c}
 \{T^{\{a}_\mu,T^{b\}}_\nu\}=U^{\{a\rho}_{\mu\nu}T^{b\}}_\rho,
\\[9pt]
\{T^a_\mu,T_\alpha\}=\tilde{U}^{a\beta}_{\mu\alpha}T_\beta+
U^\nu_{\mu\alpha}T^a_\nu,
 \\[9pt]
 \{T_\alpha,T_\beta\}=\tilde{U}^\gamma_{\alpha\beta}T_\gamma+
{1\over2}\varepsilon_{ab}W^{\mu\nu}_{\alpha\beta}(T^b_\nu\delta^\rho_\mu-
T^b_\mu\delta^\rho_\nu(-1)^{\varepsilon_\mu\varepsilon_\nu})T^a_\rho,
\\[9pt]
 \{H,T^a_\mu\}=V^\nu_\mu T^a_\nu,
\\[9pt]
 \{H,T_\alpha\}=\tilde{V}^\beta_\alpha T_\beta+
{1\over2}\varepsilon_{ab}W^{\mu\nu}_\alpha(T^b_\nu\delta^\rho_\mu-
T^b_\mu\delta^\rho_\nu(-1)^{\varepsilon_\mu\varepsilon_\nu})T^a_\rho, \\[9pt]
\{ T^{\{a}_\mu,\tilde{U}^{b\}\beta}_{\nu\alpha}\}-
\{T^{\{a}_\nu,\tilde{U}^{b\}\beta}_{\mu\alpha}\}
(-1)^{\varepsilon_\mu\varepsilon_\nu}-
\tilde{U}^{\{a\gamma}_{\mu\alpha}\tilde{U}^{b\}\beta}_{\nu\gamma}
(-1)^{\varepsilon_\nu(\tilde{\varepsilon}_\alpha+ \tilde{\varepsilon}_\gamma)}
+
\\[9pt]
+\tilde{U}^{\{a\gamma}_{\nu\alpha}\tilde{U}^{b\}\beta}_{\mu\gamma}
(-1)^{\varepsilon_\mu(\tilde{\varepsilon}_\alpha+\tilde{\varepsilon}_\gamma+
\varepsilon_\nu)}-U^{\{a\rho}_{\mu\nu}\tilde{U}^{b\}\beta}_{\rho\alpha}=
\tilde{U}^{\{a\rho\beta}_{\mu\nu\alpha}T^{b\}}_\rho
(-1)^{\varepsilon_\mu\tilde{\varepsilon}_\alpha} \, .
\end{array}
\eqno{(5)}$$

We suppose the constraints $T^{a}_{\mu}$, $T_{\alpha}$ to guarantee the
existence of structure functions (4) to satisfy the relations (5) just called
the split involution algebra.

One should remember that the involution
relations (5) define the structure functions (4) with a natural
ambiguity. This ambiguity is related to the those parts of
functions (4) which vanish on the constraint surface and do not
contribute to the constraint algebra (5).

An antisymmetric in $a,b$ part of the matrix of mutual
Poisson brackets of the constraints $T^{a}_{\mu}$ with themselves
$$
\{ T^{[a}_\mu \, , \, T^{b]}_\nu \} = 2 \varepsilon^{ab} \Delta_{\mu\nu}
$$
is not determined by the constraint algebra (2.5).
We suppose $ \Delta_{\mu\nu}$
to be an invertible matrix, which means, in its own turn, that the constraints
$T^{a}_{\mu}$ are of the second class. Besides, we suppose the complete set of
the constraints $T^{a}_{\mu}$, $T_{\alpha}$,
as well as its each subset, to be irreducible, which implies that
the Jacobi matrix
$$
{\frac{D( \, T^a_\mu \, , \, T_\alpha \, )}{D ( \, q^i \, , \, p_i \, )}
\,\,}_{
\displaystyle{\bigg |_{\displaystyle{T^a_\mu = T_\alpha = 0}}}}
$$
has the maximal rank.

Thus the problem is to prove the existence of a solution to the equations (3)
and (1), under the assumption that the constraints generate the split
involution algebra (5), and, then, to describe the characteristic
arbitrariness of the general solution.

As it has already been said in Introduction, we begin with the abelianization
of the algebra (5) with the help of the admissible transformations. With this
purpose, let us write down the admissible transformations of the constraints
and the Hamiltonian $H$ generated by the automorphisms (2).
We demonstrate explicitly only the part of the lowest orders in
ghosts in the transformation generators which may contribute to the
transformation of $H,\,T_\alpha, \, T_\mu^a$ and some essential
structure functions of the algebra (5):
$$
\begin{array}{c}
U = e^{\hat{u}}\, , \quad {\hat{u}}f=\lbrace f \, , \, u \rbrace,
\\[9pt]
u= F(q,p) + C^{\prime\prime\mu}S^\nu_\mu\CBPii_\nu(-1)^{\varepsilon_\nu}
+ C^{\prime\alpha}\tilde{S}^\beta_\alpha\CBPi_\beta
(-1)^{\tilde{\varepsilon}_\beta}+
\\[9pt]
C^{\prime\prime\nu}C^{\prime\prime\mu}S_{\mu\nu}^{\lambda\rho}
\CBPii_\rho\CBPii_\lambda(-1)^{\varepsilon_\nu+\varepsilon_\rho} +
C^{\prime\beta}C^{\prime\alpha}\tilde{S}^{\gamma\delta}_{\alpha\beta}
\CBPi_\delta\CBPi_\gamma
(-1)^{\tilde{\varepsilon}_\beta + \tilde{\varepsilon}_\delta } +
\\[9pt]
C^{\prime\alpha}C^{\prime\prime\mu}\tilde{S}^{\nu\beta}_{\mu\alpha}
\CBPi_\beta\CBPii_\nu(-1)^{\varepsilon_\mu + \tilde{\varepsilon}_\beta}
+
\frac{1}{2}
C^{\prime\alpha}C^{\prime\prime\mu}C^{\prime\prime\rho}
\tilde{S}^{\nu\lambda\beta}_{\mu\rho\alpha}
\CBPi_\beta\CBPii_\nu\CBPii_\lambda
(-1)^{\varepsilon_\mu + \tilde{\varepsilon}_\beta} +
\ldots,
\\[9pt]
\Xi ={1\over2}C^{\prime\alpha}
\Sigma^{\mu\nu}_\alpha\CBPii_\nu\CBPii_\mu
(-1)^{\varepsilon_\nu}+
\\[9pt]
{1\over6}C^{\prime\alpha}C^{\prime\prime\lambda}
\Sigma^{\mu\nu\rho}_{\alpha\lambda}
\CBPii_\rho\CBPii_\nu\CBPii_\mu
(-1)^{\varepsilon_\nu+\varepsilon_\alpha}+
{1\over2}C^{\prime\alpha}C^{\prime\beta}\Sigma^{\mu\nu\gamma}_{\alpha\beta}
\CBPi_\gamma\CBPii_\nu\CBPii_\mu
(-1)^{\varepsilon_\nu+\varepsilon_\alpha}+\ldots, \\[9pt]
\Psi= \zeta^\alpha \CBPi_\alpha + C^{\prime\alpha}\Psi^{\beta\gamma}_{\alpha}
\CBPi_\gamma\CBPi_\beta(-1)^{\tilde{\varepsilon}_\beta +
\tilde{\varepsilon}_\alpha } + C^{\prime\prime\mu}\Psi^{\beta\nu}_\mu
\CBPii_\nu\CBPi_\beta(-1)^{\varepsilon_\mu +
\tilde{\varepsilon}_\beta}+\ldots, \\[9pt]
\Phi = \frac{1}{2}\Phi^{\nu\mu}\CBPii_\nu\CBPii_\mu (-1)^{\varepsilon_\nu}+
 \\[9pt]
C^{\prime\beta}\Phi^{\nu\mu\alpha}_\beta
\CBPii_\nu\CBPii_\mu\CBPi_\alpha
(-1)^{\varepsilon_\nu + \tilde{\varepsilon}_\beta} +
C^{\prime\prime\rho}\Phi^{\nu\mu\lambda}_\rho
\CBPii_\lambda\CBPii_\nu\CBPii_\mu
(-1)^{\varepsilon_\rho + \tilde{\varepsilon}_\lambda} +
\ldots \, ,
\\[9pt]
X_a=C^{\prime\beta}C^{\prime\alpha}X_{a\alpha\beta}^{\mu\nu\lambda}
\CBPii_\lambda\CBPii_\nu\CBPii_\mu
(-1)^{\varepsilon_\mu + \tilde{\varepsilon}_\alpha} +
\ldots \, ,
\\[9pt]
Y_a=C^{\prime\alpha}Y_{a\alpha}^{\mu\nu\lambda}
\CBPii_\lambda\CBPii_\nu\CBPii_\mu
(-1)^{\varepsilon_\mu + \tilde{\varepsilon}_\alpha} +
\ldots \, .
\end{array}
\eqno{(6)}
$$
The function $F(q,p)$ in $u$ generates canonical
transformation in the original phase space, $\tilde{S}_\alpha^\beta$
and $S_\mu^\nu$ describe rotation of the first-- and second--class
constraint basis respectively.  $\Phi^{\nu\mu} $ and
$\Sigma^{\mu\nu}_\alpha$ are responsible for the second order second--class
constraint contributions to the Hamiltonian $H$ and first--class
constraints respectively, $\zeta^\alpha$ gives the coefficient at
the first--class constraint contribution to $H$:
$$
T^a_\mu \quad\rightarrow\quad \tilde{T}^a_\mu =
\left({\exp{S}}\right)_\mu^\nu
T^b_\nu(q(q^\prime,p^\prime),p(q^\prime,p^\prime)),
$$

$$
T_\alpha\quad\rightarrow\quad \tilde{T}_\alpha =
\left({\exp{\tilde{S}}}\right){}_\alpha^\beta
T_\beta(q(q^\prime,p^\prime),p(q^\prime,p^\prime))+
{1\over2}\varepsilon_{ab}\Sigma^{\mu\nu}_\alpha
(T^b_\nu\delta^\rho_\mu - T^b_\mu\delta^\rho_\nu
(-1)^{\varepsilon_\mu\varepsilon_\nu})T^a_\rho,
$$

$$
H\quad\rightarrow\quad\tilde{H} =
H(q(q^\prime,p^\prime),p(q^\prime,p^\prime))+
\zeta^\alpha T_\alpha +
{1\over2}\varepsilon_{ab}\Phi^{\mu\nu}(T^b_\nu\delta^\rho_\mu-
T^b_\mu\delta^\rho_\nu(-1)^{\varepsilon_\mu\varepsilon_\nu})T^a_\rho.
$$
The structure functions (4) related to these constraints and Hamiltonian
are transformed {\it inhomogeneously}, that just allows to abelianize their
algebra. The explicit form of the transformation
is rather cumbersome for these functions
and we omit it, although it could be restored
from relations (2),(6) in the same way as the transformation for the
constraints and Hamiltonian.

\section  {Abelianization of second--class constraints}

As the constraints $ T^{1}_{\mu}$ are irreducible ones,
they can be represented in the form
$$
T^1_\mu = \Lambda_\mu^\nu  \, \Phi_\nu, \quad
\Phi_\nu = \xi_\nu - \varphi_\nu,
$$
where $ \xi_{\nu}$ are some of canonical co-ordinates $\xi$,
$\varphi_{\nu}$ do not depend
on $\xi_{\nu}$, ${\Lambda_{\mu}}^{\nu}$ is an invertible matrix.
It follows from the constraint algebra that the $\Phi_{\nu}$ commute among
themselves, and, thus, one can choose them to be a set of new canonical
variables $q_{\nu}$. By making use of the admissible transformation
$U$ (with an appropriate matrix $S^\mu_\nu $ in relations (2.6)), which
results in a canonical one for the variables $p^{i}$, $q_{i}$, and rotates
the constraints $T^{a}_{\mu}$ by applying the matrix
${\Lambda^{-1}}_{\mu}^{\nu}$,
we make the constraints $T^{1}_{\mu}$ to take the form
$$
T^1_\mu = q_\mu \, .
$$
As $q_{\mu}$ together with  $T^{2}_{\mu}$ form a set of second-class constraints,
the $T^{2}_{\mu}$ are solvable for $p^{\mu}$:
$$
T^2_\mu = \Lambda_{\mu\nu} \Psi^{\nu} \, , \quad
\Psi^{\nu} = p^\nu  - \psi^\nu \, ,
$$
where $\psi^{\nu}$ does not depend on $p^{\mu}$.
It follows from the constraint algebra that $\Psi^{\nu}$ commute among
themselves, so that the $q_{\mu}$, $\Psi^{\mu}$ commute canonically, and,
thus, one can choose the $\Psi^{\mu}$ to be a set of canonical momenta (which
we will denote further by just $p^{\mu}$). Assuming that we have already
applied the mentioned canonical transformation, we get the following structure
for second-class constraints
$$
T^1_\mu = q_\mu \, , \quad T^2_\mu = \Lambda_{\mu\nu} p^\nu =
\lambda_{\mu\nu} p^\nu  + O ({\eta}^2) \,  ,
\eqno{(1)}
$$
where $\lambda_{\mu\nu}$ does not depend on $\eta, \,  \eta = (q_{\mu},p^{\mu})$.
The split involution relations,  which determine the Poisson brackets of
second--class constraints for $a = 1, b = 2$, yield for $\eta = 0$
$$
\lambda_{\mu\nu}  = (-1)^{\varepsilon_\mu \varepsilon_\nu}
\lambda_{\nu\mu}  \, .
$$
Recall that the $\lambda_{\mu\nu}$ is an invertible matrix. In general, the
matrix $\lambda_{\mu\nu}$ depends on the variables complement to $\eta $, and
the constraints $T^{2}_{\mu}$ are nonabelian even in linear approximation in
$\eta$. For the sake of technical simplicity we restrict ourselves in further
evaluation by the extra assumption that the only nonzero blocks of the
matrix $\lambda_{\mu\nu}$ are (symmetric) Bose-Bose block and (antisymmetric)
Fermi-Fermi one ( i.e. Bosonic and Fermionic constraints are not mixed in
linear approximation in $\eta$). It will be shown in Appendix A
that such a matrix can be represented in the form
$$
\lambda = X^T t X  \,  ,
$$
where $X$ is an invertible purely Bosonic matrix (which depends, in general, on
the variables $\xi$), and $t$ is a constant matrix whose Bose-Bose block is a
diagonal matrix with diagonal elements equal to $+1,-1$, while its Fermi-Fermi
block has a block-diagonal structure with matrices $\sigma$ being the Jordan
blocks,
$$
\sigma = \imath \sigma^2 =
\bigg(
\begin{array}{rr}
 0 & 1 \\
-1& 0
\end{array}
 \bigg),
$$
(Bose-Fermi blocks of the matrix t equal to zero).
Let us rotate simultaneously the constraints $T^{1}$ and $T^{2}$ by applying the
matrix $(X^{T})^{-1}$, which results for the constraints in their taking the form
$$
T^1 = (X^T)^{-1} q \, ,  \quad T^2 = t  X  p.
$$
Further, let us apply the canonical transformation with W being a generating function

$$
W= p^{\prime} (X^T (\xi^{\prime}_p , \xi_q))^{-1} q
$$
so that in an explicit form the transformation reads
$$
q=(X^T (\xi^{\prime}_p , \xi_q))q^{\prime} + O (q^{\prime}) p^{\prime},\quad
p=(X^T (\xi^{\prime}_p , \xi_q))^{-1}p^{\prime} + O (q^{\prime}) p^{\prime}.
$$
In new variables (we omit primes), the constraints $T^{1}$ and $T^{2}$
take the form
$$
T^1 = \Lambda^1_{\mu\nu} q_{\nu} \, , \quad \Lambda^1_{\mu\nu} =
\delta_{\mu\nu} + O(p),
$$
$$
T^2 = \Lambda^2_{\mu\nu} p_{\nu} \, , \quad \Lambda^2_{\mu\nu} =
t_{\mu\nu} + O(q).
$$
Now, let us rotate simultaneously these constraints by applying the matrix
$(\Lambda^{1})^{-1}$. It is convenient to formulate the result as saying that
the constraints $T^{1}$ and $T^{2}$ are reduced to take the form
$$
T^1_{\mu}=q_{\mu} \, , \quad T^2_{\mu}=\Lambda_{\mu\nu} p^{\nu}=
t_{\mu\nu} p^{\nu} + O(\eta^k) \, , \eqno{(2)}
$$
$$
U^{a\rho}_{\mu\nu} = \stackrel{(k - 2)}{U}{}^{a\rho}_{\mu\nu}+
O(\eta^k) \, , \quad
{\stackrel{(k - 2)}{U}}{}^{a\rho}_{\mu\nu} \sim
(\eta^{k-2}), \eqno{(3)}
$$
for $k = 2$.

Let us assume that we have already reduced the constraints $T^{1}$ and $T^{2}$
to take the form (2), (3) for $k = n > 2$. We are intended to show that, under
the above assumption, one can reduce them to take the form (2),(3) for
$k = n + 1$.

The constraint $T^{2}$ has the form
$$
T^2_{\mu}= {\bar{p}}_\mu + t_{\nu\mu} \sum\limits_{k=0}^{n-1}
C^k_{\nu | (\lambda)_k | (\sigma)_{n-k}} (q_{\lambda})^k (\bar{p}_\sigma)^{n-k}
+ O(\eta^{n+1})  \, , \quad \bar{p}_\mu \equiv t_{\mu\nu}p^\nu,
\eqno{(4)}
$$
where coefficients $C^{k}$ can depend on the variables $\xi$. It is shown in
Appendix B that the following representation holds
$$
t_{\nu\mu} \sum\limits_{k=0}^{n-1}
C^k_{\nu | (\lambda)_k | (\sigma)_{n-k}}
(q_{\lambda})^k ({\bar{p}}_{\sigma})^{n-k} =
t_{\nu\mu} \sum\limits_{k}^{n-1}
{\bar{C}}^k_{\nu |(\lambda)_k  (\sigma)_{n-k}} (q_{\lambda})^k (\bar{p}_\sigma)^{n-k}
+ {\stackrel{n-1}{a}}_{\mu | [\sigma\rho]} q_\sigma \, ,
\eqno{(5)}
$$
$$
{\stackrel{n-1}{a}}_{\mu | [\sigma\rho]} = - (-1)^{\varepsilon_\sigma\varepsilon_\rho}
{\stackrel{n-1}{a}}_{\mu | [\rho\sigma]} \, ,  \quad
{\stackrel{n-1}{a}}_{\mu | [\sigma\rho]} =  O(\eta^{n-1}) \, ,
$$
with some coefficients ${\bar{C}}^{k}_{\mu|(\lambda)_{n}}$. By making use of
simultaneous rotation of the constraints by applying the matrix
$$
{\rm exp} (-\stackrel{n-1}{a}) \, , \quad \stackrel{n-1}{a}_{\mu\nu}=
\stackrel{n-1}{a}_{\mu | [\sigma\nu]} q_\sigma,
$$
we get the constraints $T^{1}$ in the form (2) again, while the constraints
$T^{2}$ take the form (4) with ${\bar{C}}^{k}$ standing for $C^{k}$. Note that
the property (3) of the structure coefficients remains stable under these
transformations.

Let us return to the constraint algebra (2.5). For $a = b = 1$ we have
$$
U^{1\rho}_{\mu\nu} q_\rho = 0 \, , \quad U^{1\rho}_{\mu\nu} =
m^{1 | [\sigma\rho]}_{\mu\nu}q_\sigma \, .
$$
These structure coefficients can be annulled with the
help of an admissible transformation
U (2.6) (with an appropriate function $S^{\lambda\rho}_{\mu\nu}$) ,
so that we can assume that
$$
\stackrel{n-2}{U}{}^{1\rho}_{\mu\nu}
=U^{1\rho}_{\mu\nu}=0.
$$
In the same way, it follows from the algebra
relations (2.5) for $a = b = 2$ that
$$
\begin{array}{c}
\stackrel{n-2}{U}{}^{2\rho}_{\mu\nu} =
t_{{\mu}^{\prime}\mu}t_{{\nu}^{\prime}\nu} \sum\limits_{k=0}^{n-1} k
\bigg( (-1)^{\varepsilon_{\mu^\prime}\varepsilon_{\nu^\prime}}
{\bar{C}}^k_{\mu^{\prime} |\nu^{\prime} (\lambda)_{k-1}
(\sigma)_{n-k-1}\rho} - \\[9pt]
{\bar{C}}^k_{\nu^{\prime} |\mu^{\prime} (\lambda)_{k-1}  (\sigma)_{n-k-1}\rho}
 \bigg)
(q_\lambda)^{k-1}(\bar{p}_\sigma)^{n-k-1} +
\stackrel{n-3}{m}{}^{2[\sigma\rho]}_{\mu\nu} \bar{p}_\sigma \, ,
\end{array}
$$
$$
\stackrel{n-3}{m}{}^{2[\sigma\rho]}_{\mu\nu} \sim\eta^{n-3}.
$$
Finally, the algebra relations (2.5) for $a = 1, b = 2$ yield
$$
\begin{array}{c}
n  t_{{\mu}^{\prime}\mu}t_{{\nu}^{\prime}\nu}
\sum\limits_{k=0}^{n-1}
\left(
(-1)^{\varepsilon_{\mu^\prime}\varepsilon_{\nu^\prime}}
{C}^k_{\mu^{\prime} |\nu^{\prime} (\lambda)_{k-1}  (\sigma)_{n-k-1}} -
{C}^k_{\nu^{\prime} |\mu^{\prime} (\lambda)_{k-1}  (\sigma)_{n-k-1}}
 \right)
(q_\lambda)^{k}({p}^{\prime}_\sigma)^{n-k-1}
= \\
-
\stackrel{n-3}{m}{}^{2[\sigma\rho]}_{\mu\nu}\bar{p}_\sigma q_\rho.
\end{array}
\eqno{(6)}
$$
By setting $q_{\mu} = {\bar{p}}_{\mu}$ in relations (6), we get
$$
C^k_{\mu_1 | \mu_2 (\mu)_{n-1}} = (-1)^{\varepsilon_{\mu^1}\varepsilon_{\mu^2}}
C^k_{\mu_2 | \mu_1 (\mu)_{n-1}} = { s}^k_{(\mu)_{n+1}} \, ,
$$
i.e. the coefficients $C^{k}$ are totally (super)symmetric in all their indices.

Let us apply the following canonical transformation to all the variables
$$
\xi \,  \to  \,  e^{\hat{\varphi}}\xi \, , \quad {\hat{\varphi}} F =
\lbrace F \, , \, {\varphi} \rbrace  \, ,\quad
\varphi = \frac{1}{n+1} \sum\limits_{k=0}^{n-1}
s^k_{(\lambda)_{k+1} (\sigma)_{n-k}},
$$
and then rotate the constraints by applying the matrix
$$
(e^{- \stackrel{n}{b}})_{\mu\rho} \, , \quad {\stackrel{n}{b}}_{\mu\rho} =
\sum\limits_{k=0}^{n-1} \frac{n-k}{n+1} t_{\nu\mu}
s^k_{\nu(\lambda)_k(\sigma)_{n-k-1}}(q_\lambda)^k(\bar{p})^{n-k-1}.
$$
As a result, the constraints take the form
$$
T^1_\mu = \Lambda^1_{\mu\nu} q_{\nu} \, , \quad \Lambda^1_{\mu\nu} =
\delta_{\mu\nu} + O (\eta^n) ,
$$
$$
T^2_\mu = \Lambda^2_{\mu\nu} {\bar{p}}_{\nu} \, , \quad \Lambda^2_{\mu\nu} =
\delta_{\mu\nu} + O (\eta^n).
$$
Applying then the matrix $(\Lambda^{1})^{-1}$, we reduce the constraints
$T^{1}$ and $T^{2}$ to take the form (2) with $k = n + 1$.
All the transformations, which we have made use of, do not affect the property
(3) ( for $k = n$ ) of the structure coefficients.

Let us consider once more the consequences of the constraint algebra (2.5) for
them. The algebra relations (2.5) for $a = b = 1$ yield
$$
\stackrel{(n-2)}{U}{}^{1\rho}_{\mu\nu} =
\stackrel{(n-3)}{m}{}^{1[\sigma\rho]}_{\mu\nu} q_\sigma \, ,
$$
which makes it possible to annul ${\stackrel{(n - 2)}{U}}{}^{1\rho}_{\mu\nu}$
with the help of an admissible transformation
(with an appropriate function  $S^{\lambda\rho}_{\mu\nu}$ in $u$).
The algebra relations (2.5) for $a = b = 2$ yield
$$
\stackrel{(n-2)}{U}{}^{2\rho}_{\mu\nu} =
\stackrel{(n-3)}{m}{}^{2[\sigma\rho]}_{\mu\nu} {\bar{p}}^\sigma \, ,
$$
and for $a = 1$, $b = 2$ the algebra relations (2.5) imply
$$
\stackrel{(n-3)}{m}{}^{2[\sigma\rho]}_{\mu\nu}{\bar{p}}^\sigma q^\rho = 0.
\eqno{(7)}
$$
It is shown in Appendix B that the condition (7), in its own turn, implies
$$
\stackrel{(n-3)}{m}{}^{2[\sigma\rho]}_{\mu\nu} =
U^{[\lambda\sigma\rho]} q_\lambda + V^{[\lambda\sigma\rho]}
\bar{p}_\lambda
\eqno{(8)}
$$
so that
$$
\stackrel{(n-2)}{U}{}^{2\rho}_{\mu\nu}=
U^{[\lambda\sigma\rho]} q_\lambda{\bar{p}}_\sigma.
$$
Thus, the
structure coefficients $\stackrel{(n - 2)}{U}{}^{a\rho}_{\mu\nu}$ are
rewritten in the form
$$
\stackrel{(n-2)}{U}{}^{a\rho}_{\mu\nu}=
\kappa^{\sigma\rho}_{\mu\nu}T^a_\sigma + O(\eta^{n-1}) \, , \quad
\kappa^{\sigma\rho}_{\mu\nu}=U^{[\lambda\sigma\rho]}_{\mu\nu}q_\lambda
\, ,
$$
which can be annulled with the help of an admissible transformation
U (2.6) (with an appropriate function $S^{\lambda\rho}_{\mu\nu}$).
By making use of the induction method, we conclude that the second-class
constraints can be reduced to take the abelian form
$$
T^1 = q_\mu \, , \quad T^1 = t_{\mu\nu}p_\nu \equiv {\bar{p}}_\mu \, ,
\quad U^{a\rho}_{\mu\nu} = 0,
$$
with the help of the admissible transformations.

\section  {Abelianization of first--class constraints}

As the complete set of constraints is supposed to be irreducible, the
first--class constraints $T_{\alpha}$ are solvable for some variables
$\P_{\alpha}$ complement to $\eta$
$$
T_\beta = \Lambda_{\alpha\beta} (
{\P}_\alpha + t_{0|\alpha} + t^\mu_{11|\alpha}q_\mu +  t^{\mu}_{12|\alpha}
p_\mu + t_{2 |\alpha}) \, ,
$$
where $t_{0\alpha}$, $t^{\mu}_{11|\alpha}$ and
$t^{\mu}_{12|\alpha}$ do not depend on $\P$ and $\eta$, while
$t_{2|\alpha} = O({\eta}^{2})$ do not depend on $\P$. By substituting
this representation into the relation (2.5) of the constraint
algebra, we find $$ \begin{array}{c} \lbrace \P_\alpha + t_{0|\alpha}
\, , \, \P_\beta + t_{0|\beta} \rbrace = 0 \, , \\[6pt]
 t^{\mu}_{11|\alpha} = t^{\mu}_{12|\alpha} = 0 \, .
\end{array}
\eqno{(1)}
$$
These relations mean that one can choose the functions
$\P_{\alpha} +t_{0|\alpha}$
to be a set of new canonical co-ordinates (which we denote
again by just $\P_{\alpha})$. By solving the constraints $T_{\alpha}$ for
these new variables, and then rotating the constraints $T_{\alpha}$ with
the help of the corresponding matrix $\Lambda^{-1}$, we reduce the constraints
to take the form
$$
T_\alpha = \P_\alpha + t_\alpha,\quad
\frac{\partial t_\alpha }{\partial \P_\beta} = 0, \quad
t_\alpha =O(\eta^2).
$$
In accordance with Appendix B, the functions $ t_{\alpha}$ can be represented,
as for their $\eta$-dependence, in the form
$$
t_\alpha(\eta) = {s}_\alpha (q, \bar{p}) +
t_\alpha^{[\mu\nu]} q_\mu {\bar{p}}_\nu ={s}_\alpha (q, \bar{p}) +
t_\alpha^{[\mu\nu]} T^1_\mu T^2_\nu.
\eqno{(2)}
$$
The second term in (2) can be compensated by an admissible transformation
(2.6), with an appropriate structure coefficient
$\Sigma^{\mu\nu}_\alpha$ in the generator $\Xi$,
so that the constraints $T_{\alpha}$ take the form
$$
T_\alpha = \P_\alpha + {s}_\alpha (q, \bar{p}) \, , \quad
{s}_\alpha =o(\eta^2) \, , \, \, \, \frac{\partial}{\partial\P_\alpha}
{s}_\beta = 0
\eqno{(3)}
$$
(of course, ${s}_{\alpha}$ can depend on the co-ordinates conjugate to
$\P_{\alpha}$, and on physical variables as well).

Let us substitute the representation (3)
for $T_{\alpha}$ into the relation (2.5) of the constraint algebra
$$
t_{\lambda\mu} \frac{\partial}{\partial p_\lambda} {s}_\alpha =
{\tilde{U}}^{1\beta}_{\mu\alpha}T_\beta + U^{\nu}_{\mu\alpha}q_\nu,\quad
 - t_{\lambda\mu} \frac{\partial}{\partial q_\lambda} {s}_\alpha =
{\tilde{U}}^{2\beta}_{\mu\alpha}T_\beta + U^{\nu}_{\mu\alpha}{\bar{p}}_\nu.
\eqno{(4)}
$$
Next, let us set $\P_{\alpha} = 0,\, {\bar{p}}_{\mu} = q_{\mu}$ in (4), and then
subtract the second equation from the first one
$$
t_{\lambda\mu} \left(\frac{\partial}{\partial p_\lambda} {s}_\alpha +
\frac{\partial}{\partial q_\lambda} {s}_\alpha \right) \bigg|_{\bar{p}=q}
= \left( {\tilde{U}}^{1\beta}_{\mu\alpha} - {\tilde{U}}^{2\beta}_{\mu\alpha}\right)
{s}_\beta \bigg|_{\bar{p}=q, \, \P=0}.
$$
The only solution to this equation, expandable in power series in $\eta$, is
$$
{ s }_\alpha = 0 \, ,
$$
where we have taken into account that $ s_{\alpha} = O \, ({\eta}^{2})$.
Now, the relation (4) of the constraint algebra is rewritten as
$$
{\tilde{U}}^{1\beta}_{\mu\alpha} \P_\beta + U^\nu_{\mu\alpha} T^a_\nu = 0 \, .
$$
Let us represent $U^{\nu}_{\mu\alpha}$ in the form
$$
U^{\nu}_{\mu\alpha} = \stackrel{(0)}{U}{}^{\nu}_{\mu\alpha} +
U^{\nu\beta}_{\mu\alpha} {\P}_\beta \, , \quad
\frac{\partial}{\partial \P_\beta}\stackrel{(0)}{U}{}^{\nu}_{\mu\alpha} = 0.
$$
Then we get
$$
\stackrel{(0)}{U}{}^{\nu}_{\mu\alpha} q_\nu =
\stackrel{(0)}{U}{}^{\nu}_{\mu\alpha} \bar{p}_\nu =  0, \quad
({\tilde{U}}^{a\beta}_{\mu\alpha} +
U^{\nu\beta}_{\mu\alpha} T^a_\nu) \P_\beta =  0 \, .
$$
The general solution to the set of equation is (for solving of the first
equation see Appendix B)
$$
{\tilde{U}}^{a\beta}_{\mu\alpha} = -  U^{\nu\beta}_{\mu\alpha} T^a_\nu +
{\tilde{U}}^{a [\delta\beta]}_{\mu\alpha} \P_\delta,\quad
U^{\nu}_{\mu\alpha} = U^{\nu\beta}_{\mu\alpha} \P_\beta +
U^{[\lambda\sigma\nu]}_{\mu\alpha}q_\lambda \bar{p}_\sigma.
$$
The contributions of the coefficients $ U^{\nu\beta}_{\mu\alpha}$ and
$U^{[\lambda\sigma\nu]}_{\mu\alpha}$ can be compensated by the admissible
transformation (2.6) (choosing respective structure functions
$\tilde{S}^{\nu\beta}_{\mu\alpha}$
in $u$ and $\Sigma^{\mu\nu\rho}_{\alpha\lambda}$ in $\Xi$ )
which do not change the quantities
$ M \equiv \left( T^a_\mu, U^{a\rho}_{\mu\nu}, T_\alpha \right) $.
Thus we can assume that
$$
U^{\nu}_{\mu\alpha} = 0 \, , \quad
{\tilde{U}}^{a\beta}_{\mu\alpha} = {\tilde{U}}^{a[\delta\beta]}_{\mu\alpha}
\P_\delta \equiv t_{\nu\mu}\bar{U}^{a\beta}_{\nu\alpha}.
$$
Let us consider the contributions to $ {\tilde{U}}^{a\beta}_{\mu\alpha}$ to the
zeroth and first order in $\eta$, which we further denote by

$$
\stackrel{[1]}{\tilde{U}}{}^{a\beta}_{\mu\alpha} =
t_{\nu\mu}\stackrel{[1]}{\bar{U}}{}^{a\beta}_{\nu\alpha} =
\stackrel{[1]}{\tilde{U}}{}^{a [\delta\beta]}_{\mu\alpha} \P_\delta.
$$
It follows from the relations (4) of the constraint algebra that
$$
\frac{\partial}{\partial\bar{p}_\mu}\stackrel{[1]}{\bar{U}}
{}^{1\beta}_{\nu\alpha} -
(-1)^{\varepsilon_\mu\varepsilon_\nu}
\frac{\partial}{\partial\bar{p}_\nu}
\stackrel{[1]}{\bar{U}}{}^{1\beta}_{\mu\alpha} = 0 \, ,
\quad
\frac{\partial}{\partial q_\mu}
\stackrel{[1]}{\bar{U}}{}^{2\beta}_{\nu\alpha} -
(-1)^{\varepsilon_\mu\varepsilon_\nu}
\frac{\partial}{\partial q_\nu}
\stackrel{[1]}{\bar{U}}{}^{2\beta}_{\mu\alpha} = 0 \, ,
\eqno{(5)}
$$
$$
\frac{\partial}{\partial\bar{p}_\mu}
\stackrel{[1]}{\bar{U}}{}^{2\beta}_{\nu\alpha} -
(-1)^{\varepsilon_\mu\varepsilon_\nu}
\frac{\partial}{\partial\bar{p}_\nu}\stackrel{[1]}{\bar{U}}
{}^{2\beta}_{\mu\alpha} = 0 \, ,
\quad
\frac{\partial}{\partial q_\mu}\stackrel{[1]}{\bar{U}}{}^{1\beta}_{\nu\alpha} -
(-1)^{\varepsilon_\mu\varepsilon_\nu}
\frac{\partial}{\partial q_\nu}\stackrel{[1]}{\bar{U}}{}^{1\beta}_{\mu\alpha}
= 0 \, .
\eqno{(6)}
$$
The equation (5) yields
$$
\stackrel{[1]}{\bar{U}}{}^{1\beta}_{\mu\alpha} =
\frac{\partial}{\partial\bar{p}_\mu}
\stackrel{[2]}{\varphi}{}^\beta_\alpha \, , \quad
\stackrel{[1]}{\bar{U}}{}^{1\beta}_{\mu\alpha} =
\frac{\partial}{\partial q_\mu}\left( -
\stackrel{[2]}{\varphi}{}^\beta_\alpha +
\stackrel{[2]}{\psi}{}^\beta_\alpha \right)
\, ,
$$
where we can assume
$$
\stackrel{[2]}{\varphi}{}^\beta_\alpha \bigg|_{\bar{p}=0} =
\stackrel{[2]}{\psi}{}^\beta_\alpha \bigg|_{q=0} = 0,
$$
while the equation (6) yields
$$
\frac{\partial}{\partial q_\mu}\frac{\partial}{\partial\bar{p}_\nu}
\stackrel{[2]}{\psi}{}^\beta_\alpha -
(-1)^{\varepsilon_\mu\varepsilon_\nu}
\frac{\partial}{\partial q_\nu}\frac{\partial}{\partial\bar{p}_\mu}
\stackrel{[2]}{\psi}{}^\beta_\alpha = 0 \, ,
$$
or, equivalently,
$$
\frac{\partial}{\partial\bar{p}_\mu}
\stackrel{[2]}{\psi}{}^\beta_\alpha = \frac{\partial}{\partial q_\mu}
\stackrel{[2]}{\lambda}{}^\beta_\alpha,
\eqno{(7)}
$$
with some functions $\stackrel{[2]}{\lambda}{}^\beta_\alpha $.
It follows from the equation (7) that the coefficients of power series
expansion of ${\psi}^{[2]\beta}_{\alpha}$ in $q_{\mu}$ and ${\bar{p}}_{\mu}$
are totally symmetric in all indices. The functions
${\stackrel{[2]}{\phi}}{}^{\beta}_{\alpha}$ and
${\stackrel{[2]}{\psi}}{}^{\beta}_{\alpha}$
inherit the structure of the
functions ${\tilde{U}}^{a\beta}_{\mu\alpha}$ with respect to the variables
$\P_\alpha$.
Let us consider, for example, the functions
$\stackrel{[2]}{\varphi}{}^{\alpha}_{\beta}$.
It follows from the results of Appendix B that
$$
\stackrel{[2]}{\varphi}{}^{\beta}_{\alpha} =
\stackrel{[2]}{\varphi}{}^{[\delta\beta]}_{\alpha}\P_\delta +
\sum\limits_{n=0}
\stackrel{[2]}{s}{}^n_{\alpha | (\delta)_n\beta} (\P_\delta)^n.
$$
By differentiating this relation with respect to ${\bar{p}}_{\mu}$, and
multiplying then by ${\P}_{\beta}$ from the right, we get
$$
\frac{\partial}{\partial \bar{p}_\mu}
\sum\limits_{n}
\stackrel{[2]}{s}{}^n_{\alpha | (\delta)_{n+1}} (\P_\delta)^{n + 1} =0
\, \, \, \Rightarrow \, \, \,  \stackrel{[2]}{s}{}^n_{\alpha |
(\delta)_{n+1}} = 0.
$$
The functions
${\stackrel{[2]}{\psi}}{}^{\beta}_{\alpha}$ have similar structure
$$
{\stackrel{[2]}{\psi}}{}^{\beta}_{\alpha} =
{\stackrel{[2]}{\psi}}{}^{[\delta\beta]}_{\alpha}{\P_\delta}.
$$
Finally, we obtain the following representation for
$\stackrel{[1]}{\tilde{U}}{}^{a\beta}_{\mu\alpha}$:
$$
\stackrel{[1]}{\tilde{U}}{}^{a\beta}_{\mu\alpha}= \lbrace T_\mu^a \, ,\,
\stackrel{[2]}{U}{}^{\beta}_{\alpha} \rbrace +
\stackrel{[0]}{U}{}^{\nu\beta}_{\mu\alpha}T_\nu^a \, ,
$$
$$
\stackrel{[2]}{U}{}^{\beta}_{\alpha} =
\left[ \stackrel{[2]}{\varphi_1}{}^{[\delta\beta]}_\alpha -
\frac{n_q}{n_q + n_{\bar{p}}} \stackrel{[2]}{\psi_1}{}^{[\delta\beta]}_\alpha
\right] \P_\delta,\quad
\stackrel{[0]}{U}{}^{\nu\beta}_{\mu\alpha} = t_{\mu\kappa}
\frac{\partial}{\partial q_\kappa} \left( \frac{n_q}{n_q +
n_{\bar{p}}}\stackrel{[2]}{\psi_1}{}^{[\delta\beta]}_\alpha \P_\delta
\right) \frac{\stackrel{\leftarrow}{\partial}}{\partial \bar{p}_\nu},
$$
$$ n_q = q_\mu  \frac{\partial}{\partial q_\mu} \, , \quad
n_{\bar{p}} = \bar{p}_\mu  \frac{\partial}{\partial \bar{p}_\mu}.
$$
It is easy to see that one can eliminate the
$\stackrel{[1]}{U}{}^{a\beta}_{\mu\alpha}$
with the help of the admissible transformation (2.6)
(structure functions $\tilde{S}^\beta_\alpha$,
$\tilde{S}^{\nu\beta}_{\mu\alpha}$ in $u$)
which preserves the functions
entering M (which now contains $U^{\nu}_{\mu\alpha}$ as well).

So, we have ${\tilde{U}}^{a\beta}_{\mu\alpha} = O \, ({\eta}^{2})$.
Let us assume
that we have obtained, with the help of admissible transformations, the
representation
$$
{\tilde{U}}{}^{a\beta}_{\mu\alpha} =
\stackrel{(n)}{\tilde{U}}{}^{a\beta}_{\mu\alpha} + O(\eta^{n+1}),
\quad
\stackrel{(n)}{\tilde{U}}{}^{a\beta}_{\mu\alpha} \equiv t_{\nu\mu}
\stackrel{(n)}{\bar{U}}{}^{a\beta}_{\nu\alpha} =
 t_{\nu\mu}
\stackrel{(n)}{\bar{U}}{}^{a[\delta\beta]}_{\nu\alpha}\P_\delta \sim \eta^n,
 \, \, \, n \ge 2,
$$
$$
{\tilde{U}}^{a\beta\rho}_{\mu\nu\alpha} =
\stackrel{(n-2)}{\tilde{U}}{}^{a\beta\rho}_{\mu\nu\alpha} + O(\eta^{n-1}),
\quad
\stackrel{(n-2)}{\tilde{U}}{}^{a\beta\rho}_{\mu\nu\alpha} \equiv
t_{\nu^{\prime}\nu} t_{\mu^{\prime}\mu}
\stackrel{(n-2)}{\bar{U}}{}^{a\beta\rho}_{\mu^{\prime}\nu^{\prime}\alpha}
\sim \eta^{n-2}.
$$
It follows from the relation (2.5) of the constraint algebra for $ a = b = 1$
that
$$
\frac{\partial}{\partial \bar{p}_\mu}
\stackrel{(n)}{\bar{U}}{}^{1[\delta\beta]}_{\nu\alpha}\P_\delta
- (-1)^{\varepsilon_\mu\varepsilon_\nu}
\frac{\partial}{\partial \bar{p}_\nu}
\stackrel{(n)}{\bar{U}}{}^{a[\delta\beta]}_{\mu\alpha}\P_\delta =
\stackrel{(n-2)}{\bar{U}}{}^{1\beta\rho}_{\mu\nu\alpha} q_\rho.
\eqno{(8)}
$$
The general solution to these equations is
$$
\stackrel{(n)}{\tilde{U}}{}^{1\beta}_{\mu\alpha} = \lbrace q_\mu \, , \,
\stackrel{[n+1]}{\varphi}{}_\alpha^{[\delta\beta]} \P_\delta \rbrace +
\stackrel{[n-1]}{\varphi}{}_{\mu\alpha}^{\lambda[\delta\beta]}
\P_\delta q_\lambda,
$$
$$
\stackrel{(n-2)}{\tilde{U}}{}^{1\beta\rho}_{\mu\nu\alpha}=
 \lbrace q_\mu \, , \,
\stackrel{[n-1]}{\varphi}{}_{\nu\alpha}^{\rho{[\delta\beta]}}
\P_\delta \rbrace
- (-1)^{\varepsilon_\mu\varepsilon_\nu}  \lbrace q_\nu \, , \,
\stackrel{[n-1]}{\varphi}{}_{\mu\alpha}^{\rho{[\delta\beta]}}
\P_\delta \rbrace +
\stackrel{(n-3)}{\tilde{U}}{}^{1\beta[\lambda\rho]}_{\mu\nu\alpha}q_\lambda.
$$
Because of their structure, these functions can be
compensated with the help of an admissible transformation
(structure functions $\tilde{S}^\beta_\alpha$,
$\tilde{S}^{\nu\beta}_{\mu\alpha}$ in $u$), so that
one can assume
$\stackrel{(n)}{\tilde{U}}{}^{1\beta}_{\mu\alpha}=
\stackrel{(n-2)}{\tilde{U}}{}^{1\beta\rho}_{\mu\nu\alpha} = 0 $.
Let us represent the functions
$\stackrel{(n)}{\tilde{U}}{}^{2\beta}_{\mu\alpha}$ in
the form
$$
\stackrel{(n)}{\tilde{U}}{}^{2\beta}_{\mu\alpha} =
\stackrel{(n)}{\tilde{U}}{}^{2[\delta\beta][\nu\lambda]}_{\mu\alpha}
\P_\delta q_\nu \bar{p}_\lambda +
t_{\mu^{\prime}\mu}
\stackrel{(n)}{V}{}_{\mu^\prime\alpha}^{[\delta\beta]}\P_\delta \, ,
\eqno{(9)}
$$
where the functions $\stackrel{(n)}{V}{}_{\mu^\prime\alpha}^{[\delta\beta]}$
have the structure
$$
\stackrel{(n)}{V}{}_{\mu\alpha}^{[\delta\beta]} =
\sum\limits_k {C}_{\mu\alpha|(\lambda)_k
(\sigma)_{n-k}}^{[\delta\beta]}
(q_\lambda)^k (\bar{p}_\sigma)^{n-k}.
$$
The first term in (9) can be
eliminated by the transformation (2.2), (2.6) (with appropriate
structure functions $\tilde{S}^{\nu\beta}_{\mu\alpha}$ in $u$),
which leads to the appearance of function
$\stackrel{(n-2)}{\tilde{U}}{}^{1\beta\lambda}_{\mu\nu\alpha}$
of the form
$$
\stackrel{(n-2)}{\tilde{U}}{}^{1\beta\rho}_{\mu\nu\alpha} =
\lbrace q_{[\mu} \, , \,
\stackrel{(n)}{\tilde{U}}{}^{[\delta\beta][\lambda\rho]}_{\nu]\alpha}
\P_\delta \rbrace q_\lambda
$$
which, in its own turn, can be eliminated by the admissible transformation
(2.2), (2.6) where the essential contribution is made by structure
function $\tilde{S}^{\nu\lambda\beta}_{\mu\rho\alpha}$ in $u$.

Further, the relations (2.5) of the constraint algebra for $a=1$, $b=2 $
and $a = b = 2$ yield
$$
\frac{\partial}{\partial{\bar{p}}_\mu}
\stackrel{(n)}{V}_{\nu\alpha}^{[\delta\beta]}\P_\delta
- (-1)^{\varepsilon_\mu \varepsilon_\nu}
\frac{\partial}{\partial{\bar{p}}_\nu}
\stackrel{(n)}{V}_{\mu\alpha}^{[\delta\beta]}\P_\delta  =
\stackrel{(n-2)}{\bar{U}}{}^{2\beta\rho}_{\mu\nu\alpha} \bar{p}_\rho,
\eqno{(10)}
$$
$$
\frac{\partial}{\partial {q}_\mu}
\stackrel{(n)}{V}_{\nu\alpha}^{[\delta\beta]}\P_\delta
- (-1)^{\varepsilon_\mu \varepsilon_\nu}
\frac{\partial}{\partial q_\nu}
\stackrel{(n)}{V}_{\mu\alpha}^{[\delta\beta]}\P_\delta  =
- \stackrel{(n-2)}{\bar{U}}{}^{2\beta\rho}_{\mu\nu\alpha} q_\rho.
\eqno{(11)}
$$
By setting $q_\rho = \bar{p}_\rho $ in (10)  and (11), and then
summing up them, we get the equation which implies the functions
$ C_{\mu\alpha|(\lambda)_n}^{k[\delta\beta]}$ to be totally symmetric
in the indices $\mu , \, (\lambda)_n$.

Then the equations (10), (11) yield
$$
\stackrel{(n-2)}{\tilde{U}}{}^{2\beta\rho}_{\mu\nu}T^a_\rho = 0.
$$
Finally, we find
$$
\stackrel{(n)}{\tilde{U}}{}^{2\beta}_{\mu\alpha} =
t_{\nu\mu}\frac{\partial}{\partial q_\nu}
\stackrel{(n+1)}{\psi}{}^{[\delta\beta]}_{\alpha} \P_\delta \, ,
\eqno{(12)}
$$
$$
\stackrel{(n-2)}{\tilde{U}}{}^{2\beta\rho}_{\mu\nu\alpha} =
\stackrel{(n-4)}{V}{}^{\beta[\lambda\sigma\rho]}_{\mu\nu\alpha}
q_\lambda \bar{p}_\sigma.
$$
The set of functions
$\stackrel{(n)}{\tilde{U}}{}^{1\beta}_{\mu\alpha} = 0 $,
$\stackrel{(n)}{\tilde{U}}{}^{2\beta}_{\mu\alpha}$ is
represented in the form
$$
\stackrel{(n)}{\tilde{U}}{}^{a\beta}_{\mu\alpha} =
\lbrace T^a_\mu \, , \,
\stackrel{(n+1)}{U}{}^{\beta}_{\alpha} \rbrace +
\stackrel{(n-1)}{U}{}^{\nu\beta}_{\mu\alpha} T^a_\nu \, ,
\eqno{(13)}
$$
$$
\stackrel{(n+1)}{U}{}^{\beta}_{\alpha}= - \frac{n_q}{n+1}
\stackrel{(n+1)}{\psi}{}^{[\delta\beta]}_{\alpha} \P_\delta
\, , \, \,  \,
\stackrel{(n-1)}{U}{}^{\nu\beta}_{\mu\alpha} =  \frac{1}{n+1}
t_{\lambda\mu} \frac{\partial}{\partial q_\lambda}
\stackrel{(n+1)}{\psi}{}^{[\delta\beta]}_{\alpha} \P_\delta
 \frac{\vec{\partial}}{\partial \bar{p}_\nu} \, .
$$
By making use of an admissible transformation (2.2), (2.6), one
eliminates the functions (12), (13)
$\stackrel{(n-4)}{V}{}^{\beta[\lambda\sigma\rho]}_{\mu\nu\alpha},$
$\stackrel{(n+1)}{U}{}^\alpha_\beta$,
$\stackrel{(n-1)}{U}{}^{\nu\beta}_{\mu\alpha}$
by structure functions (see (2.6))
$\tilde{S}^{\nu\lambda\beta}_{\mu\rho\alpha}$,
$\tilde{S}^{\alpha}_{\beta}$, $\tilde{S}^{\nu\beta}_{\mu\alpha}$ in $u$,
respectively.
Thus we obtain
$$
\stackrel{(n)}{\tilde{U}}{}^{a\beta}_{\mu\alpha} = 0
\, , \quad
\stackrel{(n-2)}{\tilde{U}}{}^{a\beta\rho}_{\mu\nu\alpha} = 0.
$$
As a concluding step, the relations (2.5) of the constraint algebra yield
$$
\tilde{U}^\gamma_{\alpha\beta} \P_\gamma = 2
W^{[\mu\nu]}_{\alpha\beta} q_\mu \bar{p}_\nu.
$$
By representing $W^{[\mu\nu]}_{\alpha\beta} $ in the form
$$
W^{[\mu\nu]}_{\alpha\beta} = \omega^{[\mu\nu]}_{\alpha\beta} +
W^{\gamma [\mu\nu]}_{\alpha\beta} \P_\gamma \, , \quad
\frac{\partial}{\partial \P_\gamma} \omega^{[\mu\nu]}_{\alpha\beta}
=0 \, ,
$$
we find the equations
$$
\omega^{[\mu\nu]}_{\alpha\beta}q_\mu \bar{p}_\nu = 0 \, , \quad
\left( \tilde{U}^\gamma_{\alpha\beta} -
2W^{\gamma [\mu\nu]}_{\alpha\beta} q_\mu \bar{p}_\nu \right)\P_\gamma
= 0 \, ,
$$
whose general solution is
$$
\tilde{U}^\gamma_{\alpha\beta} =
2W^{\gamma [\mu\nu]}_{\alpha\beta} q_\mu \bar{p}_\nu +
\tilde{U}^{[\delta\gamma]}_{\alpha\beta}\P_\delta \,  ,
$$
$$
W^{[\mu\nu]}_{\alpha\beta}=W^{\gamma [\mu\nu]}_{\alpha\beta}\P_\gamma
+ \omega^{[\sigma\mu\nu]}_{1 \alpha\beta} q_\sigma  +
\omega^{[\sigma\mu\nu]}_{2 \alpha\beta} \bar{p}_\sigma  \, .
$$
These structures can be completely eliminated with the help of
admissible transformations (2.6):
$W^{\gamma[\mu\nu]}_{\alpha\beta}$,
$\tilde{U}^{[\delta\gamma]}_{\alpha\beta}$,
$\omega^{[\sigma\mu\nu]}_{a \alpha\beta}$
are compensated by $\Sigma^{\mu\nu\gamma}_{\alpha\beta}$,
$\tilde{S}^{\delta\gamma}_{\alpha\beta}$,
$X^{[\sigma\mu\nu]}_{a \alpha\beta}$
respectively.

So, by making use of the admissible transformations, we have completely
abelianized the constraint algebra
$$
T^1_\mu = q_\mu \, , \quad T^2_\mu = t_{\mu\nu}p_\nu \, , \quad
T_\alpha = \P_\alpha \, ,
$$
and all the structure coefficients vanish.

\section  {Abelianization of Hamiltonian}

Let us represent the Hamiltonian $H$ in the form
$$
H = H_s + H^{[\mu\nu]} q_{\mu}\bar{p}_{\nu} \, ,
\eqno{(1)}
$$
where coefficients of power series expansion of $H_s$ with respect to
$q_{\mu}$ and ${\bar{p}}_{\mu}$ are totally symmetric in all indices. The
second term in r.h.s. of the representation (1) can be eliminated with
the help of an admissible transformation ($H^{[\mu\nu]}$ is
compensated by  $\Phi^{\mu\nu}$ entering $\Phi$ (2.6)).
The relation (2.5) of the constraint algebra yields
$$
t_{\lambda\mu}\frac{\partial}{\partial \bar{p}_\lambda} H_s =
V_\mu^\nu q_\nu \, , \quad t_{\lambda\mu}\frac{\partial}{\partial
q_\lambda} H_s = - V_\mu^\nu \bar{p}_\nu. \eqno{(2)}
$$
By summing up the equations (2), and setting then $\bar{p}_\mu = q_\mu$
, we find that $H_s$ is quite independent of  $ q_\mu$  and
$\bar{p}_\mu$. As a consequence,
$$
V_\mu^\nu T_\nu^a = 0.
$$
The general solution to this equation is
$$
V_\mu^\nu  =
V_\mu^{[\lambda\sigma\nu]}q_\lambda \bar{p}_\sigma
$$
which structure
can be eliminated with the help of an admissible transformation
$\Phi_\mu^{[\lambda\sigma\nu]}$ (2.6).

Let us represent $H_{s}$ in the form
$$
H_s =H_0 + H^\alpha \P_\alpha \, ,
\eqno{(3)}
$$
where $H_{0}$ does not depend on   $\P_\alpha$ . The second term in
r.h.s.  of (3) can be eliminated with the help of an admissible
transformation $\zeta^\alpha$ (2.6).  The relations (2.5) of the
constraint algebra yield
$$
\lbrace H_0 \, , \, \P_\alpha \rbrace =
\tilde{V}_\alpha^\beta \P_\beta - 2 W^{\mu\nu}_\alpha q_\nu
\bar{p}_\mu.
$$
It follows hereof that $H_{0}$ depends only on the
canonical variables $\omega$ complement to  $q_\mu , \, \bar{p}_\mu ,
\, Q_\alpha , \, \P_\alpha$  ($Q_\alpha$ are variables canonically
conjugate to $\P_\alpha$):
$$
  H_0=H_{ph} (\omega) \, ,
$$
and,
hence, the relations hold
$$
\tilde{V}_\alpha^\beta \P_\beta - 2
W^{\mu\nu}_\alpha q_\nu \bar{p}_\mu = 0.
$$
The general solution to these equations is
$$
\tilde{V}_\alpha^\beta = 2
W^{\beta[\mu\nu]}_\alpha q_\nu \bar{p}_\mu +
\tilde{V}_\alpha^{[\delta\beta]} \P_\delta \, ,
$$
$$
W^{\mu\nu}_\alpha = W^{\beta[\mu\nu]}_\alpha \P_\beta +
\omega_{1\alpha}^{[\sigma\mu\nu]}q_\sigma +
\omega_{2\alpha}^{[\sigma\mu\nu]}\bar{p}_\sigma \, .
$$
These structure coefficients can be eliminated with the
help of an admissible transformation (2.6) with appropriate
functions $\Phi^{\beta[\mu\nu]}_\alpha$, $Y_{a\alpha}^{\sigma\mu\nu}$.

So, we conclude that the split involution constraint algebra can be made
completely abelian in the form
$$
T^1_\mu = q_\mu \, , \quad T^2_mu = t_{\mu\nu}p_\nu \, , \quad
T_\alpha = \P_\alpha \, , \quad H = H_{ph} (\omega) \, ,
$$
by making use of the admissible transformations, and all the structure
coefficients of the algebra vanish.

\section  { Structure of general solution}

First of all, let us note that a solution to the generating equations
certainly exists. Indeed, by making use of the natural automorphism
transformations, we can reduce the constraint algebra to take the abelian
form. In the case, an obvious solution of the canonical structure does
exist
$$
\Omega^a \equiv \Omega^a_0 = C^{\prime\prime\mu}T^a_\mu \, , \quad
\Omega \equiv \Omega_0 = C^{\prime\alpha}T_\alpha \, , \quad
\CH \equiv \CH_0 = H_{ph} (\omega) \,,
\eqno{(1)}
$$
$$
K=\Lambda=0.
$$
By applying the inverse of the automorphism transformation, we obtain a
solution to the generating equations, corresponding to the original
constraint algebra.

Now, let us describe the structure of the general solution to the generating
equations, which we seek for in the form of power series expansions with
respect to the ghost momenta $ {\CBP}^{\prime}_\alpha , \,
\CBP^{\prime\prime}_\mu$:
$$
\Omega^a = \Omega^a_0 + \sum\limits_{n=1} \Omega_n^a \, , \quad
 \Omega_n^a \sim  \sum\limits_{k=0}^n C^{\prime\prime k+1}
 C^{\prime n-k} \CBP^{\prime n-k}\CBP^{\prime\prime k},
$$
$$
\Omega = \Omega_0 +  \sum\limits_{n=1} \Omega_n \, , \quad
 \Omega_n \sim  \sum\limits_{k=0}^n C^{\prime\prime k}
 C^{\prime n-k+1} \CBP^{\prime n-k}\CBP^{\prime\prime k},
$$
$$
\CH = \CH_0 +  \sum\limits_{n=0} \CH_n \, , \quad
 \CH_n \sim  \sum\limits_{k=0}^n C^{\prime\prime k}
 C^{\prime n-k} \CBP^{\prime n-k}\CBP^{\prime\prime k},
$$
$$
K =  \sum\limits_{n=1} K_n \, , \quad
 K_n \sim  \sum\limits_{k=0}^{n-1} C^{\prime\prime k}
 C^{\prime n-k+1} \CBP^{\prime n-k-1}\CBP^{\prime\prime k+2},
$$
$$
\Lambda =  \sum\limits_{n=1} \Lambda_n \, , \quad
 \Lambda_n \sim  \sum\limits_{k=0}^{n-1} C^{\prime\prime k}
 C^{\prime n-k} \CBP^{\prime n-k-1}\CBP^{\prime\prime k+2}.
$$
where we have taken into account the
conservation of $gh'$ and $gh''$.
We further assume that we have
applied the automorphism transformation which abelianizes the
constraint algebra and preserves $gh'$ and $gh''$, so that the
expressions for $\Omega^a_0 , \, \Omega_0$  and $\CH_0$ are
given by the formulae (1) , and
$$
\Omega^a_1 = \Omega_1 = \CH_1 = K_1 =\Lambda_1 =0 \, .
$$
Let us assume we have checked that any solution can be reduced, with the
help of an automorphism transformation preserving gh$^\prime$ and
gh$^{\prime^\prime}$'', to take the form   
$$
\Omega^a = \Omega^a_0 +
\Omega^a_{m} + \Omega^a_{m+1}+ O\left(\CBP^{\prime m-k+2}, \CBP^{\prime\prime
k} \right) \, ,
$$
$$ \Omega = \Omega_0 + \Omega_{m} + \Omega_{m+1}+
\left(\CBP^{\prime m-k+2}, \CBP^{\prime\prime k} \right)  \, ,
$$
$$
\CH = \CH_0 + \CH_{m} + \CH_{m+1}+ O\left(\CBP^{\prime m-k+2},
\CBP^{\prime\prime k} \right)  \, ,
$$
$$
K = K_{m} + O\left(\CBP^{\prime
m-k}, \CBP^{\prime\prime k} \right)  \, , \quad \Lambda = \Lambda_{m} +
O\left(\CBP^{\prime m-k} \, , \CBP^{\prime\prime k} \right)  \, , \quad  m
\geq  1 .
$$
It follows from the generating equations (2.3) that
$$
W^a\Omega^b_m  + W^b\Omega^a_m  = 0 \, , \quad W^a \equiv T^a_\mu
\frac{\partial}{\partial \CBP^{\prime\prime}_\mu} \, ,
\eqno{(2)}
$$
$$
(-1)^{\varepsilon_\mu}\CBP^{\prime\prime}_\mu  \left[ \lbrace T^a_\mu
\, , \, \Omega^b_m \rbrace +  \lbrace T^b_\mu \, , \, \Omega^a_m
\rbrace \right] +
W^a\Omega^b_{m+1}  + W^b\Omega^a_{m+1}  = 0 \,.
\eqno{(3)} $$
The general solution to the equations (2) is obtained in Appendix B
in the following form:
$$
\Omega^a_m =W^a X_m + \Omega^{\prime a}_m \, , \quad
\Omega^{\prime a}_m \sim C^{\prime\prime\mu} Z^a_{m\mu} , \quad
Z^a_{m\mu} \sim \left( C^\prime \CBP^{\prime} \right)^m \, ,
$$
 $Z^a_{m\mu}$  do not depend on $\CBP^{\prime\prime}_\mu$, and, thus,
 on $C^{\prime\prime\mu}$. By making use of the automorphism
transformation (with $u=X_m$), which eliminates the contribution
$W^a X_m $, we come to $\Omega^a_m = \Omega^{\prime a}_m $ , and then
it follows from the equation (3) that

$$
 \lbrace T^a_\mu
\, , \, Z^b_{m\mu} \rbrace +  \lbrace T^b_\mu \, , \, Z^a_{m\mu}
\rbrace  + (-1)^{\varepsilon_\mu \varepsilon_\nu}(\mu \leftrightarrow
\nu ) = Z^{a\lambda}_{\mu\nu}T^b_\lambda +
Z^{b\lambda}_{\mu\nu}T^a_\lambda \, ,
$$
where $ Z^{a\lambda}_{\mu\nu}$  do not depend on
$\CBP^{\prime\prime}_\mu$, $C^{\prime\prime\mu}$, and are
determined by the condition
$$
\Omega^a_{m+1}\frac{\stackrel{\leftarrow}{\partial}}{\partial
\CBP^{\prime\prime}_\lambda} \bigg|_{\CBP^{\prime\prime}=0} =
- \frac{1}{2} C^{\prime\prime\mu}C^{\prime\prime\nu}
(-1)^{\varepsilon_\mu + \varepsilon_\mu \varepsilon_\nu}
Z^{a\lambda}_{\mu\nu} \, .
$$
These equations have exactly the same structure as the equations for
$\stackrel{(n)}{\tilde{U}}{}^{a\beta}_{\mu\alpha}$, which have been
solved in Section 4.  By repeating word by word the same
reasoning, one shows the $\Omega_m^{\prime a}$  to have the structure
which can be eliminated with the help of the automorphism transformations.

The generating equations (2.3) yield
$$
W^a \Omega_m = 0 \, ,
\eqno{(4)}
$$
$$
(-1)^{\varepsilon_\mu} C^{\prime\prime \mu}\lbrace T_\mu^a \, , \,
\Omega_m \rbrace + W^a \Omega_{m+1} = 0.
\eqno{(5)}
$$
The general solution is
$$
\Omega_m = W^2W^1 \, Y_m + \Omega^{\prime}_m \, ,
$$
where $ \Omega^{\prime}_m $ does not depend on  $C^{\prime\prime \mu}$
and $\CBP^{\prime\prime}_{\mu}$, and
the coefficients of its power series expansion with respect to
$q_\mu$ and $\bar{p}_\mu$  are totally symmetric
in all indices. By eliminating the contribution $ W^2W^1 \, Y_m $ to
$\Omega_m$ with the help of the automorphism transformation $\Xi=Y_m$
(2.6), we derive form the equations (5) that $\Omega^\prime_m$ is
quite independent of $q_\mu$ and $\bar{p}_\mu$.

Now, the generating equations (2.3) yield
$$
W\Omega^\prime_m=0 \, , \quad
W =T_\alpha \frac{\partial}{\partial \CBP^{\prime}_{\alpha}},
\eqno{(6)}
$$
$$
W^2 W^1 K_m = 0.
\eqno{(7)}
$$
The general solution to these equations is
$$
\Omega^\prime_m=W Y^\prime_m \, , \quad
\frac{\partial}{\partial C^{\prime\prime}} Y^\prime_m =
\frac{\partial}{\partial \CBP^{\prime\prime}} Y^\prime_m =
\frac{\partial}{\partial T^a} Y^\prime_m = 0 \, , \quad
K_m= W^1Z_{1m} + W^2Z_{2m}.
$$
These structures can be eliminated with the help of the automorphism
transformations (2.3) generated by $X_a$.

The equations for $\CH_m$ and $\Lambda_m$, which follow from the
generating equations (2.3), coincide exactly with the equations
(6), (7) for $\Omega_m$ and $K_m$, and their solution has the
same structure which can
be eliminated with the help of the automorphism transformations.

By applying the induction method, we come to the following results:
i) a solution to the generating equations (2.3) does exist;
ii) the characteristic arbitrariness of the general solution
is described as a set of all possible automorphism transformations,
so that, in particular, any solution can be constructed by applying
an automorphism transformation to the canonical solution (1).

{\bf Acknowledgements}

The work is supported by the INTAS--RFBR grant no 95-0829. The work of I.A.B.
is partially supported by RFBR grant 96-01-00482. The work of S.L.L. is
partially supported by RFBR grant 98-02-16261. The work of I.V.T. is
partially supported by RFBR grant 96-02-17314.

\section*  {Appendix  A }

In this Appendix we prove the diagonalization lemma for symmetric and
antisymmetric matrices, which we have made use of in Section 3.

Let $G$ be a Grassmann algebra over the field R of real numbers, with
$\eta^{\alpha}, \, \, \alpha = 1, 2, \ldots , n,$ being a set of generatrices.
We call the elements of $G$ "real" as well. Given $ p \times p $ real
invertible matrices $S_{\mu\nu}$ and $A_{\mu\nu}$ with the properties
$$
S_{\mu\nu}= S_{\nu\mu}, \quad A_{\mu\nu}=  - A_{\nu\mu},\quad
\varepsilon \left(S_{\mu\nu}\right)=
\varepsilon \left(A_{\mu\nu}\right)=0,
$$
then there exist real invertible matrices $X_{1}$ and $X_{2}$,
$\varepsilon\left( X_{1}\right) = \varepsilon\left( X_{2}\right) = 0$,
such that
$$
\left( X^T_1 S X_1 \right)_{\mu\nu}= \theta_\mu \delta_{\mu\nu} \equiv
\Theta_{\mu\nu} \, ,
\eqno{(1)}
$$
$$
\left( X^T_1 A X_1 \right)_{\mu\nu} = \Sigma_{\mu\nu}
\, , \quad  \Sigma_{\mu\nu} \equiv diag (\sigma, \ldots , \sigma )\, ,
\eqno{(2)}
$$
where $\theta_{\mu} = +1, -1$ , and the number of positive (negative)
$\theta_{\mu}$ coincide with the number of positive (negative) eigenvalues
of the matrix $S_{0} = S(\eta = 0)$.

Proof.

i)  {\bf Matrix $S$.}

Let us expand the matrix $S$ in (finite) power series with respect to the
generatrices $\eta^{\alpha}$
$$
S = \sum\limits_{k=0} S_{2k}, \quad S_{2k} \sim (\eta)^{2k}.
$$
As the $S_{0}$ is a real symmetric matrix, it can be diagonalized with the
help of an orthogonal real transformation $O^{(0)}$. Let us consider the matrix
$$
S^{(0)}= E O^{(0)T} S O^{(0)} E,
$$
where $E_{\mu\nu} = e_{\mu} \delta_{\mu\nu}$ is an invertible matrix.
The matrix $S^{0}$ has the following structure
$$
S^{(0)}= S^{(0)}_0 + S^{(0)}_2 + O(\eta^4) \, , \quad S^{(0)}_2 \sim
\eta^2 \, ,
$$
$$
S_{0|\mu\nu} = \theta_\mu |\lambda_\mu|
e^2_\mu\delta_{\mu\nu}\equiv\ s^{(0)}\delta_{\mu\nu}\, , \quad
\theta_\mu =\frac{\lambda_\mu}{|\lambda_\mu|} \, ,
$$
where $\lambda_{\mu}$ are the eigenvalues of the matrix $S^{0}$.
The numbers $e_{\mu}$ are chosen
to satisfy the condition that there are no equal numbers among
$s^{(0)}_{\mu}$ (which can always be fulfilled).

Let us consider the matrix
$$
S^{(2)}= O^{(2)T} S^{(0)} O^{(2)}  \, ,
$$
$$
O^{(2)}_{\mu\nu}=\delta_{\mu\nu} -
\frac{S^{(0)}_{2|\mu\nu}}{s^{(0)}_\mu - s^{(0)}_\nu} \, , \,\,\,
\mu \neq \nu \, ,
\quad O^{(2)}_{\mu\mu} = 1 \, \, ( {\rm no \, \, sum \, \, over}.
\mu)
$$
Its $\eta$-power series expansion has the form
$$
S^{(2)}= S^{(2)}_0 + S^{(2)}_4 + O(\eta^6) \, , \quad S^{(2)}_4 \sim
(\eta)^4,
$$
$$
S^{(2)}_{0|\mu\nu} = s^{(2)}_\mu \delta_{\mu\nu} \, ,  \quad
s^{(2)}_\mu= s^{(0)}_\mu S^{(0)}_{2|\mu\mu} \equiv
\theta_\mu \kappa^{(2)}_\mu \, , \quad
\kappa^{(2)}_\mu=|\lambda_\mu| e^2_\mu + \theta_\mu
S^{(0)}_{2|\mu\mu}.
$$
Further, let us introduce the matrix $S^{(4)}$
$$
S^{(4)}=O^{(4)T}S^{(2)}O^{(4)} \, ,
$$
$$
O^{(4)}_{\mu\nu}=\delta_{\mu\nu} -
\frac{S^{(2)}_{4|\mu\nu}}{s^{(0)}_\mu - s^{(0)}_\nu} \, , \,\,\,
\mu\neq\nu \, , \quad O^{(4)}_{\mu\mu}=1 \, ,
$$
which diagonalizes $S^{(2)}$ already in the fourth order in $\eta$.  By
continuing this process, and taking into account that $(\eta)^{n+1} = 0$,
we get at the $2m$-th step
$$
\left[\left( O^{(0)} E O^{(2)} \dots O^{(2m)} \right)^T S
\left( O^{(0)} E O^{(2)} \dots O^{(2m)} \right)\right]_{\mu\nu}=
 \kappa^{(2m)}_\mu \theta_\mu \delta_{\mu\nu} \, ,
$$
$$
\kappa^{(2m)}_\mu=|\lambda_\mu| e^2_\mu + O(\eta^2) = \sigma_{\mu}^2
\, , \quad
\sigma_{\mu}=e_\mu \sqrt{|\lambda_\mu|} + O(\eta^2).
$$
It is now easy to see that one can choose
$$
X_{1|\mu\nu}=
\left( O^{(0)} E O^{(2)} \dots O^{(2m)}
\right)_{\mu\nu}\frac{1}{\sigma_\nu}
$$
to serve as the matrix $X_{1}$.

Note that our result extends to cover the case of a Grassmann algebra over
the field $C$ of complex numbers, as follows: given a symmetric
invertible matrix $S_{\mu\nu}$, whose elements belong to $G$, and
$\varepsilon (S_{\mu\nu}) = 0$, then there exists an invertible
matrix $X_{\mu\nu}$, whose elements belong to $G$, and $\varepsilon
(X_{\mu\nu}) = 0$, such that the relation holds
$$
S=X^T \, X.
$$

ii) {\bf Matrix $A$.}

As the matrix $A$, by assumption, is invertible, p is an even number: $p = 2q$.
Let us expand the matrix $A$ in power series with respect to the generatrices
$$
A = \sum\limits_{k=0} A_k \, , \quad A_k \sim (\eta)^{2k}.
$$
As the $A_{0}$ is a real antisymmetric matrix, it can be reduced, with the help
of an orthogonal transformation $O^{(0)}$, to take the canonical form
$$
O^{(0)T}A_{0}O^{(0)}= diag (\lambda_1 \sigma, \dots , \lambda_q\sigma)
\, , \quad \lambda_l > 0 \, , \, \, \, l=1 , \ldots , q.
$$
Let us introduce the matrix
$$
A^{(0)}=E O^{(0)T}A O^{(0)} E,
$$
$$
E = diag (e_1 I_2 , \ldots \, e_q I_2 ) \, ,
$$
where $I_{2}$ is the unit $2 \times 2$ matrix.
The matrix $A^{0}$ expands in power series in $\eta$ in the form
$$
A^{(0)} = diag (a^{(0)}_1 \sigma , \ldots , a^{(0)}_q \sigma ) +
A^{(0)}_2 + O \left( (\eta)^4 \right) \, , \quad A^{(0)}_2 \sim
(\eta)^2,
$$
$$
a^{(0)}_l = \lambda_l e^2_l > 0 \, .
$$
The numbers $e_{l}$ are chosen in such a way that all the
numbers $a^{(0)}_{l}$ are different. It is further convenient to
split the indices $\mu$, $\nu$ as follows
$$
\mu = (a,l) \, , \quad \nu = (b,l) \, , \quad a,b = 1,2  \, , \quad
l,l^\prime = 1, \ldots , q \, ,
$$
$$
A^{(0)}_{2 |\mu\nu} =A^{(0)}_{2|(a,l)(b,l^\prime)} \, , \quad
A^{(0)}_{2|(a,l)(b,l)} = A^{(0)}_{2|l}\sigma \, , \quad
A^{(0)}_{2|(a,l)(b,l^\prime)}  = - A^{(0)}_{2|(b,l^\prime)(a,l)}.
$$
Let us introduce the matrix $A^{(2)}$:
$$
A^{(2)}=O^{(2)T}A^{(0)}O^{(2)} \, , \quad O^{(2)}_{\mu\nu} =
\delta_{\mu\nu}  + O^{(2)}_{2|\mu\nu},
$$
$$
O^{(2)}_{2|(a,l)(b,l)}=0 \, , \quad O^{(2)}_{2|(a,l)(b,l^\prime)}=
\frac{a^{(0)}_l\sigma_{ac} A^{(0)}_{2|(c,l)(b,l^\prime)} +
A^{(0)}_{2|(a,l)(c,l^\prime)} \sigma_{cb} a^{(0)}_{l^\prime}}{(a^{(0)}_{l})^2
- (a^{(0)}_{l^\prime})^2} \, , \, \, \, l\neq l^\prime \, ,
$$
which expands in power series in $\eta$ as
$$
A^{(2)} = diag (a^{(2)}_1 \sigma , \ldots , a^{(2)}_q \sigma ) +
A^{(2)}_4 + O \left( (\eta)^6\right) \, , \quad A^{(2)}_4 \sim (\eta)^4 \,,
$$
$$
a^{(2)}_l = a^{(0)}_l + A^{(0)}_{2|l}\,.
$$
Then we act quite in the same way as in the case of diagonalization of a
symmetric matrix, so that at the $2m$-th step we have
$$
\left( O^{(0)} E O^{(2)} \dots O^{(2m)} \right)^T A
\left( O^{(0)} E O^{(2)} \dots O^{(2m)} \right)=
diag (a^{(2m)}_1 \sigma , \ldots , a^{(2m)}_q \sigma ) \, ,
$$
$$
a^{(2m)}_l=\lambda_l e^2_l +  O(\eta^2) = \kappa_l^2 \, , \quad
\kappa_l = e_l \sqrt{|\lambda_l|} + O(\eta^2) \, .
$$
Then one can choose
$$
X_{2|\mu\nu} = \left( O^{(0)} E O^{(2)} \dots O^{(2m)}
\right)_{\mu\lambda} \left( diag \left(\frac{1}{\kappa_1}I_2, \ldots
, \frac{1}{\kappa_1}I_2\right) \right)_{\lambda\nu}
$$
to serve as the matrix $X_{2}$.

The result (2) remains valid in the case of Grassmann algebra over
the field $ C $ of complex numbers.

\section*  { Appendix  B }

In this Appendix we prove the existence lemma for special representations
of functions of two sets of variables, which we have made use of in Section 3.

i) We shall need the following well known result. Given the operator
$d$,
$$
d = x^i\frac{\partial}{\partial x^*_i} \, , \quad d^2 = 0 \, , \quad
\varepsilon (x^*_i) = \varepsilon (x^i) + 1 \equiv
 \varepsilon_i +1 \, .
$$
the general solution to the equation
$$
dX=0
$$
can be represented in the form
$$
X = d Y + C \, , \quad \frac{\partial C}{\partial x^*_i} =
\frac{\partial C}{\partial x^i} = 0 \, ,
$$
where $Y = Y (x, x^{*})$ is an arbitrary function, and $C$
cannot be represented in the form of $du$.

Let us introduce the pair of operators $d_{a}$
$$
d_1 = x^i\frac{\partial}{\partial x^*_i} \, , \quad
d_2 = y^i\frac{\partial}{\partial x^*_i} \, , \quad
\varepsilon (x^i) = \varepsilon (y^i) \, ,
$$
to form the algebra
$$
d^a d^b + d^b d^a = 0.
$$
Let us construct the general solution to the equation
$$
d_2 d_1 X = 0 \, .
\eqno{(1)}
$$
Let us expand $X$ in power series in $x$ and $y$:
$$
X=\sum\limits_{m,n}X_{m,n} \, ,  \quad X_{m,n} \sim (x)^m(y)^n.
$$
It is obvious that each of $X_{m,n}$ satisfies the equation (1):
$$
d_2 d_1 X_{m,n} = 0 \, .
$$
By making use of the above result, one can write down
$$
\begin{array}{cclrcl}
d_1X_{m\, , n\, \,}&=& d_2 Y_{m+1, n-1} \, , \quad & d_2 d_1 Y_{m+1,
n-1} & = & 0, \\
d_1Y_{m+1,n-1 }&=& d_2 Y_{m+2, n-2} \, ,
\quad & d_2 d_1 Y_{m+2, n-2} & = & 0, \\
{} & \dots & {} &{} &{} &{}\\
d_1Y_{m+n - 1 ,  1 }&=& d_2 Y_{m+ n, \, 0} \, , \quad
& d_2 d_1 Y_{m+n, \,0} & = & 0,
\end{array}
$$
$$
d_1 Y_{m+ n, \, 0} =  s^{m,n}_{(j)_{m+n+1}} (x_j)^{m+n+1} =
d_1 (x^*_is^{m,n}_{i(j)_{m+n}} (x_j)^{m+n}) \, ,
$$
where $s^{m,n}$ are constants. For $Y_{m+n,0}$ we get
$$
Y_{m+n,0}=x^*_i s^{m,n}_{i(j)_{m+n}} (x_j)^{m+n}
+ d_1Z_{m+n-1,0}.
$$
Let us substitute this expression into the equation for
$Y_{m+n-1,1}$ \,
$$
d_1 \left( Y_{m+n-1,1} - d_2 Z_{m+n-1,0} \right) =
s^{m,n}_{(j)_{m+n} l} (x_j)^{m+n} y_l =
d_1( x^*_i s_{i(j)_{m+n-1}l}^{m,n} (x_j)^{m+n-1}y_l) \, ,
$$
which yields
$$
Y_{m+n-1,1} = d_2 Z_{m+n-1,0} + d_1 Z_{m+n-2,1} + x^*_i
s_{i(j)_{m+n-1}l}^{m,n} (x_j)^{m+n-1} y_l.
$$
Finally, we obtain
$$
X_{m,n}= d_1 Z_{m-1,n} + d_2 Z_{m,n-1} + x^*_i
s_{i(j)_m(l)_n}^{m,n} (x_j)^m (y_l)^n.
\eqno{(2)}
$$
The third term in (2) cannot be represented in the form of $ d_{1} \,
F $ or $ d_{2} \, F $ because
$$
d_1\left(x^*_is_{i(j)_m(l)_n}^{m,n} (x_j)^m (y_l)^n\right) \neq 0 \, , \quad
d_2\left(x^*_is_{i(j)_m(l)_n}^{m,n} (x_j)^m (y_l)^n\right) \neq 0 \, .
$$

iii) Let us consider the set of equations
$$
d_a X = 0. \eqno{(3)}
$$
For $ a = 1$ we have
$$
X=d_1Y +s(y) \, , \quad
s(y)=\sum\limits_{n=0}s_{(l)_n}^{0,n}(y_l)^n \, .
$$
It follows from the equation (3) for $a = 2 $ that
$$
Y=d_1Z_1+d_2Z_2+
x^*_i\sum\limits_{m=1,n=0}s_{i(j)_m(l)_n}^{m,n} (x_j)^m (y_l)^n \, ,
$$
$$
X=d_2d_1Z+s(x,y)\, , \quad
s(x,y)=\sum\limits_{m,n=0}s_{(j)_m(l)_n}^{m,n} (x_j)^m (y_l)^n \, .
$$
The function $ s (x,y)$ can not be represented in the form of $d_2d_1 F$.
Indeed, let
$$
s(x,y)=d_2d_1 F.
\eqno{(4)}
$$
By setting $y_i=x_i$ in (4), we get
$$
s(x,x)=0,
$$
which implies
$s(x,y)=0$.
Finally, let us consider the set of equations
$$
d_a u_b + d_b u_a = 0 \, .
\eqno{(5)}
$$
For $a = b = 1$ we have
$$
d_1u_1=0 \, ,
$$
which implies
$$
u_1=d_1Y_1+Z_1(y) \, .
$$
The equation (5) for $a = 1, b = 2$ yields
$$
d_1u_2=d_1d_2Y_1 \, ,
$$
which implies
$$
u_2 =d_2Y_1+d_1Y_2+Z_2(y) \, .
$$
The equation (5) for $a = b = 2$ yields
$$
d_2d_1Y_2=0 \, ,
$$
which implies
$$
Y_2=d_1V_1+d_2V_2+x^*_i s_i(x,y) \, , \quad
s_i(x,y)\equiv\sum\limits_{m=1,n=0}s_{i(j)_m(l)_n}^{m,n} (x_j)^m (y_l)^n \, .
$$
Thus we have
$$
u_2=d_2Y_1-d_2d_1V_2+x_i s_i(x,y) \, .
$$
Denoting
$$
Y=Y_1-d_1V_2 \,  , \quad s_2(x,y) = x_i s_i(x,y)  - Z_2(y), \quad
s_1(y)\equiv Z_1(y) \, ,
$$
we conclude that the general solution to the equations (5) can be
represented in the form
$$
u_a=d_aY+s_a \, , \quad s_1=s_1(y)\, , \quad s_2=s_2(x,y),
\eqno{(6)}
$$
where the second term cannot be represented in the form of $d_{a} u$.

Let $\phi (x,y)$ be a given function.  It is obvious to satisfy the equations
$$
d_a\phi = 0.
$$
Then the functions $s(x,y)$ and $z = z^{ij} x^{*}_{i} x^{*}_{j}, \,
z^{ji}=(-1)^{(\varepsilon_i+1)(\varepsilon_j+1)} z^{ij}$ should exist
such that
$$
\phi = d_2d_1z+s= a^{ij}x_iy_j + s,
$$
$$
a^{ij}=2z^{ij}(-1)^{\varepsilon_i}.
$$
It is obvious that
$$
a^{ij}= -(-1)^{\varepsilon_i\varepsilon_j} a^{ji}=a^{[ij]}
$$
which implies just the representation (3.5).

Let $ m^{[ij]}$, $\varepsilon (m^{[ij]}) = \varepsilon_{m} + \varepsilon_{i}
+ \varepsilon_{j}$, be a given function to satisfy the condition
$$
m^{[ij]}x_iy_j=0.
$$
Then the function $m = m^{[ij)} (-1)^{\epsilon_{i}} x^{*}_{i} x^{*}_{j}$
satisfies the equation
$$
d_2d_1m=0
$$
and, hence,
$$
m=d_1Z_1+d_2Z_2\, , \quad Z_a=Z_a^{ijk}x^*_ix^*_jx^*_k \, ,
$$
so that we get hereof
$$
m^{[ij]}=U^{[lij]}x_l+V^{[lij]}y_l \, ,
$$
$$
U^{[lij]}=(-1)^{\varepsilon_m+\varepsilon_l+\varepsilon_j}Z_1^{lij} \,
, \quad
V^{[lij]}=(-1)^{\varepsilon_m+\varepsilon_l+\varepsilon_j}Z_2^{lij}\, ,
$$
which implies just the representation  (3.8).

Let $ m^{i}$, $\varepsilon (m^{i}) = \varepsilon_{m} + \varepsilon_{i}$, be a
given function to satisfy the condition
$$
m^ix_i=m^iy_i=0.
$$
Then the function $ m = (-1)^{\epsilon_{i}} m^{i} x^{*}_{i}$ satisfies the
equations
$$
d_am=0 \, ,
$$
so that we get hereof
$$
(-1)^{\varepsilon_i}m^ix_i^*=d_2d_1Z^{jli}x_j^*x_l^*x_i^*=
(-1)^{\varepsilon_i}m^{[jli]}x_jy_lx_i^* \, ,
$$
$$
m^{[jli]}=6Z^{jli}(-1)^{\varepsilon_j+\varepsilon_i} \, ,
$$
which implies
$$
m^i= m^{[jli]}x_jy_l.
$$

\end{document}